\documentclass[]{aa}  

\usepackage{placeins}

\usepackage{graphicx}
\usepackage{txfonts}
\usepackage{subcaption}

\usepackage[normalem]{ulem}

\usepackage{gensymb}

\usepackage[thinc]{esdiff}

\usepackage[version=4]{mhchem}
\usepackage{chemformula}

\AddToHook{begindocument/before}{\usepackage{hyperref}}

\begin{document}

   \title{Observational constraints on CO sequestration and disk mass in the planet-hosting PDS~70 disk}

   \subtitle{}

   \author{M. Leemker \inst{1}
          \and
          S. Facchini \inst{1}
          \and
          F. Zagaria \inst{2}
          \and
          K. Doi \inst{2}
          \and
          J. Bae \inst{3}
          \and 
          M. Benisty \inst{2}
          \and 
          H. B. Liu \inst{4, 5}
          \and 
          C. Longarini \inst{6}
          \and 
          V. Pezzotta \inst{1}
          \and 
          L. Rampinelli \inst{1}
          \and 
          L. M. Stapper \inst{2}
          \and 
          R. Teague \inst{7}          
          }

        \institute{Dipartimento di Fisica, Università degli Studi di Milano, Via Celoria 16, 20133 Milano, Italy\\
              \email{margot.leemker@unimi.it}
        \and
        Max Planck Institute for Astronomy, Königstuhl 17, 6911 7, Heidelberg, Germany
        \and
        Department of Astronomy, University of Florida, Gainesville, FL 32611, USA
        \and 
        Department of Physics, National Sun Yat-Sen University, No. 70, Lien-Hai Road, Kaohsiung City 80424, Taiwan, R.O.C.
        \and 
        Center of Astronomy and Gravitation, National Taiwan Normal University, Taipei 116, Taiwan
        \and
        Institute of Astronomy, University of Cambridge, Madingley Rd, Cambridge CB30HA, UK
        \and
        Department of Earth, Atmospheric, and Planetary Sciences, Massachusetts Institute of Technology, Cambridge, MA 02139, USA
         }

   \date{Received XXX; accepted XXX}

  \abstract
  {Exoplanets accrete gas, dust, and ice from their natal environment, setting the chemical composition of their atmospheres and interiors. The abundance of the major volatiles in disks can be used to possibly link exoplanet atmospheres to their formation location in the disk. In this work, we aim to quantify the peak CO abundance and disk mass of the PDS~70 disk, the first disk with a direct confirmation of two accreting planets in its dust cavity.
  We present the first detection of \ce{N2H+} and \ce{C^17O} in this disk. The \ce{C^17O} emission shows marginal statistical evidence for an azimuthal asymmetry trailing that previously seen in the continuum emission. We then compare the disk integrated fluxes of \ce{N2H+}, CO isotopologues, and the continuum to an existing grid of thermochemical models. We find that the derived disk mass drops by a factor of $\sim 3.4$ from $0.039^{+0.067}_{-0.018}$~M$_{\odot}$ to $0.011^{+0.011}_{-0.006}$~M$_{\odot}$ if the \ce{C^17O} line flux is included in the fitting procedure; the quoted intervals do not include systematic uncertainties. CO on the other hand is sequestered with an effective abundance of $(4.8^{+4.3}_{-2.7} - 17^{+8}_{-8}) \times 10^{-6}$ relative to \ce{H2} in the probed disk layers after accounting for CO freeze-out and isotope selective CO photodissociation for the adopted chemical network and ionization prescription. The range in peak CO abundances reflects the uncertainty in the \ce{N2H+} emission, assuming either optically thin emission at 21~K or optically thick emission. Therefore, the PDS~70 planets are possibly accreting the final portion of their atmospheres from the outer disk molecular layer where CO is sequestered. Thus, CO sequestering may need to be considered when interpreting the chemical composition of planetary atmospheres.
  }

   \keywords{protoplanetary disks - submillimeter: planetary systems – stars: individual: PDS 70 - astrochemistry}

   \maketitle
%

\section{Introduction}
Exoplanets that are observed to date are shaped by the mass of their natal environment and the chemical differentiation of key elements such as carbon, oxygen, and nitrogen between the gas, ice, and refractory material. Therefore, the chemical composition of exoplanetary atmospheres could hold important clues to constrain the formation location of the planets in their natal protoplanetary disk \citep[e.g.,][]{Oberg2011, Madhusudhan2017, Cridland2019, Walsh2025}. 

In protoplanetary disks, the chemical composition of gas and ices changes as a function of radius due to the freeze-out and desorption of the major volatiles at their respective snowlines \citep{Oberg2011, Oberg2023_review}. The bulk of the ice reservoir is likely inherited from the earliest phases of star and planet formation \citep{Booth2021, Andreu2023, Tobin2023, Leemker2025_V883Ori, Slavicinska2025}. These ices are oxygen rich due to the high desorption temperature of water that only desorbs in the innermost regions of disks \citep{Harsono2015, Carr2018, Bosman2021, Leemker2021, Facchini2024, Leemker2026, Rampinelli2026}. On the other hand, the gas outside the water snowline is generally carbon rich \citep[e.g.,][]{Bergin2016, Miotello2019, Bosman2021_MAPS_C_to_O}. 

The most abundant gas-phase molecule in the ISM after \ce{H2} is CO with an abundance of $10^{-4}$ relative to \ce{H2} \citep{Dickman1978}. Throughout this work the CO abundance is defined as $n(\mathrm{CO})/n(\mathrm{H_2})$ with $n$ the number density of CO and \ce{H2}, respectively. 
In recent years, observations of T~Tauri disks have shown that these disks have weaker CO emission than expected compared to the CO abundance in the ISM, whereas this is typically not the case for Herbig disks \citep[e.g.,][]{Ansdell2016, Long2017, Miotello2017, Zhang2019, Zhang2021_coco, Anderson2022, Trapman2022, Trapman2025_exoALMA, Trapman2025_agepro, Stapper2024}. This was already suggested by observations of the HD $1-0$ line, which provides a CO-independent measurement of the disk mass \citep{Bergin2013, McClure2016, Kama2016, Schwarz2016, Trapman2017, Sturm2023}. The weak CO emission in T~Tauri disks has been explained by the isotope selective photodissociation, freeze-out, locking of dust with CO ice in the disk midplane, and chemical conversion of CO \citep[e.g.,][]{Krijt2020}. During the latter process, CO is converted to larger molecules such as \ce{CO2} and complex organic molecules that remain frozen out onto the grains up to the innermost regions of the disk \citep[e.g.,][]{Bosman2018}.  

\begin{table*}
\caption{Observed molecular lines and Band~3 continuum with ALMA in the PDS~70 disk.\label{tab:det_ALMA}}
\centering
\begin{tabular}{lccccccccc}
\hline\hline
Transition & Freq. & $r_{\rm flux}$ & Int. flux$^{(a)}$ & $\Delta V_{\rm chan}$ &$\sigma_{\rm chan}$ & \texttt{robust} & $uv$-taper & Beam & $\epsilon $  \\
 & (GHz)& & (mJy km~s$^{-1}$) & (km~s$^{-1}$) & (mJy beam$^{-1}$) & &  &   &  \\
\hline
\ce{N2H+} $1-0$ &  93.173 & $1\farcs8$ & $76 \pm 14$ & 3.1 & 0.4 & 2.0 & $0\farcs2$ & $0\farcs56\times 0\farcs48\ (-75\degree)$ & 0.34  \\
\ce{C^17O} $3-2$ & 337.061 & $1\farcs8$ & $390 \pm 45$ & 1.7 & 0.6 & 2.0 & $0\farcs1$ & $0\farcs35\times 0\farcs32\ (76\degree)$ & 0.54 \\
Continuum & 97.5 & $1\farcs2$ & $4.2 \pm 0.2$$^{(b)}$ & - & 0.007 & 0.0 & - & $0\farcs098\times 0\farcs079\ (-40\degree)$ & N/A  \\
\hline
\end{tabular}
\tablefoot{$^{(a)}$ The uncertainties include the 5\% (\ce{N2H+} and continuum) and 10\% (\ce{C^17O}) absolute flux calibration uncertainty of ALMA. $^{(b)}$ The continuum flux is listed in units of mJy.}
\end{table*}

The resulting gas-phase CO abundance in disks can be calibrated using \ce{N2H+}, a molecule that is destroyed by gas-phase CO \citep{vantHoff2017, Anderson2022, Trapman2022}. Therefore, if CO is heavily sequestered, i.e. the CO abundance is lower than expected after accounting for freeze-out and isotope selective photodissociation, \ce{N2H+} is expected to be bright. The results from the recent ALMA large programs AGE-PRO \citep{Zhang2025} and exoALMA \citep{Teague2025} utilizing this method confirm earlier results demonstrating that CO is sequestered by roughly an order of magnitude compared to the ISM value \citep{Trapman2025_exoALMA, Trapman2025_agepro}.

In this work, we measure the disk mass and quantify the peak CO abundance in the PDS~70 disk, the first disk with a direct detection of two accreting planets through H$\alpha$ emission. In Sect.~\ref{sec:obs}, the detection of \ce{N2H+} and \ce{C^17O} in this disk are presented. 
We then leverage the framework developed by \citet{Trapman2025_agepro} in Sect.~\ref{sec:methods_mcmc} to constrain the disk mass and CO abundance using these line fluxes together with archival continuum and other CO isotopologue observations, enabling a direct comparison with other disks analyzed using the same framework.
The resulting CO abundance and disk mass are presented in Sect.~\ref{sec:results} and discussed in Sect.~\ref{sec:discussion}. Finally, our conclusions are summarized in Sect.~\ref{sec:conclusions}.

\section{Observations} \label{sec:obs}

In this work, we present the first detection of \ce{N2H+} and \ce{C^17O} in the PDS~70 disk. The \ce{N2H+} $J=1-0$ transition at 93.173~GHz is covered in the Band~3 spectral setup of ALMA program 2022.1.01477.S (PI: H. Liu; \citealt{Doi2024, Liu2024}) in a 1.9~GHz wide continuum spectral window (spw) centered at 92.4~GHz with a spectral resolution of 977~kHz (3.1~km~s$^{-1}$). The Band~7 setup of the same program covers the \ce{C^17O} $J=3-2$ transition at 337.061~GHz. The spw covering this line is centered at 336.5~GHz and has a spectral resolution of 1953~kHz (1.7~km~s$^{-1}$) and a total bandwidth of 1.9~GHz. Both the Band~3 and the Band~7 setup include three additional spectral windows with similar settings centered at 90.5, 102.5, and 104.5~GHz (Band~3) and 338.4, 348.5, and 350.5~GHz (Band~7). The data from program 2022.1.00893.S also cover the \ce{N2H+} line but are not included our analysis as the spectral resolution of 100~km~s$^{-1}$ is unsuited for the analysis of spectral lines.

Both the Band~3 and the Band~7 setup were observed in a compact configuration resulting in short baselines (SBs) and an extended configuration with long baselines (LBs). A summary of the observations is presented in Table~\ref{tab:obs_details}. Before combining the Execution Blocks (EBs) of the Band~3 data, a pseudo-continuum of all four spectral windows in each EB was created by flagging the data within 15~km~s$^{-1}$ of the bright lines and the \ce{N2H+} $1-0$ transition and averaging over a 30~MHz bandwidth to avoid bandwidth smearing \citep{Bridle1989}. Then a continuum image was made using the \texttt{CASA} version 6.5.4 task \texttt{tclean} with the \texttt{mtmfs} deconvolver \citep{Rau2011} with 2 terms and a central frequency of 97.5~GHz. The images were cleaned using Briggs weighting \citep{Briggs1995} with a robust parameter of 0.5 down to a 4$\sigma$ threshold. The relative alignment of these EBs was determined following the procedure outlined in \citet{Loomis2025}. Finally, these alignment offsets were applied to the non-averaged, non-flagged data before concatenating the three EBs. The continuum was subtracted from this combined Measurement Set using the task \texttt{uvcontsub\_old} that fits a first order polynomial to the line free regions in the spectrum. 

Each EB in Band~7 is self-calibrated for a single round of phase-only self-calibration using a solution interval with the length of the EB after flagging the data within 15~km~s$^{-1}$ of bright lines. Following the same procedure as for the Band~3 data, the individual EBs are aligned following the procedure outlined in \citet{Loomis2025} before concatenating all EBs. Finally, the data are continuum subtracted.

\begin{figure*}
\centering
\includegraphics[width=\hsize]{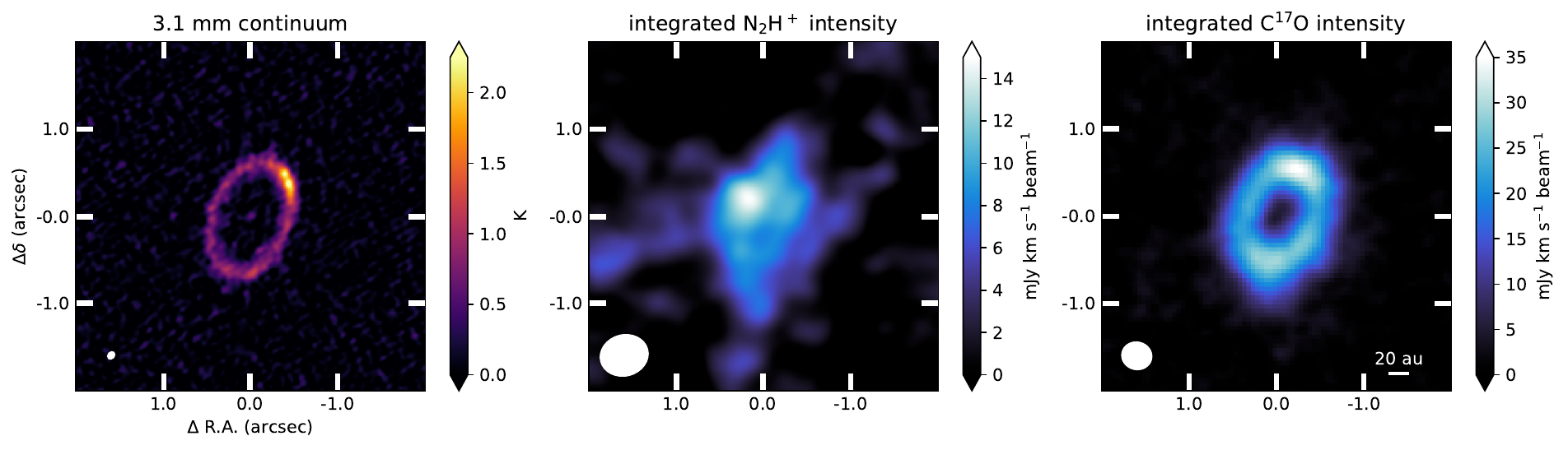}
  \caption{ALMA Band~3 and 7 observations in the PDS~70 disk. Left: 3.1~mm continuum emission. Middle: integrated intensity map of the continuum subtracted \ce{N2H+} $J=1-0$ emission (moment 0). Right: integrated intensity map of the continuum subtracted \ce{C^17O} $J=3-2$ emission. The beam is indicated in the bottom left corner of each panel and a 20~au scalebar is shown in the bottom right corner of the last panel. }
     \label{fig:mom0_N2H+}
\end{figure*}

\subsection{Imaging}

The resulting Band~3 continuum image is presented in the left panel of Fig.~\ref{fig:mom0_N2H+}. This image is created in a similar way to those used for the alignment but with a robust parameter of 0 instead of 0.5 to provide a slightly better spatial resolution of $0\farcs098\times 0\farcs079\ (-40.5\degree)$ at the cost of a somewhat higher rms of 7.4~$\mu$Jy~beam$^{-1}$. In addition, this image was cleaned down to a threshold of $1\sigma$. This is needed to include as much flux as possible in the \texttt{clean} model to accurately measure the flux by accounting for the difference between non-Gaussian shape of the beam in the dirty image and the Gaussian clean beam. 

The \ce{N2H+} and \ce{C^17O} lines were imaged at the native spectral resolution of the data using the \texttt{multiscale} deconvolver with scales [0,8,15,30,80] times the pixel size of $0\farcs015$ for the \ce{N2H+} and [0,8,15] times the pixel size of $0\farcs05$ for the \ce{C^17O} emission. During the cleaning, we used a Keplerian mask centered at the continuum and convolved with a Gaussian two (\ce{C^17O}) and three (\ce{N2H+}) times larger than the beam for a $1\farcs0$ disk with an inclination of 51.7$\degree$, a position angle of 160.4$\degree$, a stellar mass of 0.87~M$_{\odot}$, a source velocity of 5.5~km~s$^{-1}$, and a distance of 113~pc \citep{gaia2018, Keppler2019, Teague_kep_mask}. No separate masks were created for the hyperfine components as the spectral resolution of the data is too low to separate these from the main component. Similar to the continuum, the lines were cleaned down to a 1$\sigma$ threshold. In addition, the line cubes are JvM-corrected to accurately measure the flux by accounting for the difference between the non-Gaussian shape of the beam in the dirty image and the Gaussian clean beam \citep{JvM, Czekala2021}. This is particularly relevant for weak (line) emission where a large part of the flux is not captured in the clean model. To not overestimate the significance of the emission, all the noise estimates are measured using the non-JvM corrected cubes \citep{Casassus2022}. The JvM-correction factor $\epsilon$ is listed in Table~\ref{tab:det_ALMA}.

As the emission is somewhat extended, the baselines were weighted using a Briggs weighting scheme with a robust parameter of 2 and a $0\farcs1$ $uv$-taper for the \ce{C^17O} emission and a  $0\farcs2$ $uv$-taper for the \ce{N2H+} emission. The resulting \ce{C^17O} and \ce{N2H+} cubes have a $0\farcs35\times 0\farcs32\ (76\degree)$ and $0\farcs56\times 0\farcs48\ (-75\degree)$ beam and a channel rms of 0.6~mJy~beam$^{-1}$ and 0.4~mJy~beam$^{-1}$, respectively. The channel maps are presented in Fig.~\ref{fig:chans_N2H+} and ~\ref{fig:chans_C17O}.

To further analyze the \ce{N2H+} and \ce{C^17O} detections in the PDS~70 disk, the line cubes were collapsed using the better moments package \citep{Teague_better_moments}. The integrated intensity (moment~0) and the intensity weighted velocity (moment~1) maps were both computed between $-$5.3 and 16.7~km~s$^{-1}$ for the \ce{N2H+} and between 1.5 and 12.0~km~s$^{-1}$ for the \ce{C^17O}. The velocity range for the \ce{N2H+} $1-0$ emission is wider due to the lower velocity resolution of those data and the larger velocity spread between the hyperfine components of 15~km~s$^{-1}$ compared to 1.4~km~s$^{-1}$ for the \ce{C^17O} $J=3-2$ emission. The moment~1 maps use the same velocity range but a 3$\sigma$ clip was applied. The resulting integrated intensity maps are presented in the middle and right panels of Fig.~\ref{fig:mom0_N2H+} and the moment~1 maps are presented in Fig.~\ref{fig:mom1_N2H+}.

\subsection{Flux measurements}

We measure a total Band~3 continuum flux of $4.2\pm 0.2$~mJy inside a $1\farcs2\times 0\farcs74$ (semi-axes) region, which is consistent with previous measurements in ALMA Band~3 \citep{Liu2024}. A higher signal-to-noise ratio image is presented in \citet{Doi2024} who combined the data used in this project with that of ALMA project 2022.1.00893.S covering a similar spectral range in time division mode. For the lines, we measure a flux of $76 \pm 14$~mJy~km~s$^{-1}$ for the \ce{N2H+} and $390 \pm 45$~mJy~km~s$^{-1}$ for the \ce{C^17O} inside a $1\farcs8 \times 1\farcs1$ ellipse (semi-axes). The 
uncertainty is propagated from the channel rms in the non-JvM corrected cube using standard error propagation \citep[see e.g.,][for the equations used]{Leemker2026}. These uncertainties include the 5\% (Band~3) and 10\% (Band~6 and 7) absolute flux calibration uncertainty of ALMA \citep{Francis2020, Cortes2025_ALMA_technical_handbook}. We verified that the measured line flux is independent of the imaging parameters for a wide range of robust values and/ or $uv$-tapers. The resulting \ce{N2H+} and \ce{C^17O} image parameters are summarized in Table~\ref{tab:det_ALMA}.

\begin{table}
\caption{Summary of the fluxes used for the \texttt{mcmc} sampling.  \label{tab:fit_inputs}}
\centering
\begin{tabular}{lccc}
\hline\hline
Molecular transition & Freq. & $F_{\nu}\Delta V$ & Ref. \\
& (GHz) & (Jy km~s$^{-1}$) &\\
\hline
\ce{N2H+} $3-2$$^{(a)}$   & 279.512 & $2.12\pm 0.40$     & 1 \\
\ce{^12CO}        $3-2$   & 345.796 & $12.4 \pm 1.3$     & 2 \\ 
\ce{^12CO}        $2-1$   & 230.538 & $6.18 \pm 0.62$    & 2, 3  \\ 
\ce{^13CO}        $2-1$   & 220.399 & $1.86 \pm 0.19$    & 2, 3  \\ 
\ce{C^18O}        $2-1$   & 219.560 & $0.466 \pm 0.052$  & 2, 3  \\ 
\ce{C^17O}        $3-2$   & 337.061 & $0.390 \pm 0.045$  & 1 \\ 
Continuum$^{(b)}$ & 231 GHz & $0.0614 \pm 0.006$ & 2 \\
Continuum$^{(c)}$ & 273 GHz & $0.084 \pm 0.009$  & 3 \\
Continuum$^{(d)}$ & 337 GHz & $0.16 \pm 0.02$    & 4, 5 \\
\hline
\end{tabular}
\tablefoot{The uncertainties on the line fluxes include the 5\% and 10\% absolute flux calibration uncertainty of ALMA for the \ce{N2H+} and the other observations, respectively. $^{(a)}$ The flux of the $3-2$ transition is estimated from the $1-0$ by assuming LTE and an excitation temperature of 21~K. $^{(b-d)}$ Fluxes are listed in units of Jy and are scaled from the observed frequencies of 225, 253, and 351~GHz to the frequencies of 231, 273, and 337~GHz included in the model grid assuming a spectral index of 2.8$\pm 0.4$ \citep{Doi2024, Fasano2025, Sierra2025}. \textbf{References:} 1. This work; 2. \citealt{Facchini2021}; 3. \citealt{Rampinelli2024}; 4. \citealt{Isella2019}; 5. \citealt{Benisty2021}.}
\end{table}

In addition to the \ce{N2H+} and \ce{C^17O} data presented in this work, we collect the disk integrated fluxes from previously published interferometric data of the continuum and CO isotopologue emission in the PDS~70 disk. The uncertainty on the \ce{^12CO}, \ce{^13CO}, and \ce{C^18O} $J=2-1$ line flux reported in \citet{Rampinelli2024} are measured on the JvM-corrected cubes. In Table~\ref{tab:fit_inputs}, we report the uncertainties for these three measurements using the non-JvM corrected cubes to not underestimate the uncertainty and for consistency with the other flux measurements that are all estimated from the non-JvM corrected cubes.

\section{Methods} \label{sec:methods_mcmc}
To constrain the gas mass and the peak CO abundance with respect to \ce{H2} in the disk layers probed by the observations in the PDS~70 disk, we follow the methods outlined in \citet{Trapman2025_exoALMA} and \citet{Trapman2025_agepro} where the 90\% flux radius of the \ce{^13CO} emission and the observed fluxes of CO isotopologues, \ce{N2H+}, and the continuum are compared to those in a large grid of thermochemical \texttt{DALI} models \citep{Bruderer2012, Bruderer2013}. The results from this method are in good agreement with independent disk mass measurements from rotation curves for the disks in the exoALMA large program \citep{Longarini2025, Trapman2025_exoALMA}. As the PDS~70 disk has a large central dust cavity but still shows some \ce{^12CO} and \ce{HCO+} emission in the center \citep{Long2018, Facchini2021}, we use the transition disk model grid\footnote{\url{https://zenodo.org/records/15148628}} where the models have a dust cavity but no gas cavity rather than the grid with full dust disk models. We note that there is no model grid available where the disks have a cavity in both the dust and the gas. The effect of the latter is further discussed in Sect.~\ref{sec:discussion_systematics}. The models vary not only in disk mass and the peak CO  abundance with respect to \ce{H2}, but also in the radial structure (characteristic radius $R_{\mathrm{c}}$), dust size distribution, gas-to-dust mass ratio $\Delta_{\rm g/d}$, vertical structure, stellar luminosity $L_{\star}$, and the cosmic ray ionization rate $\zeta_{\rm c.r.}$ used for the \ce{N2H+} chemical network. The CO abundance is treated as a free parameter where the abundance is varied in steps of a factor of 10 in log-space after accounting for CO freeze-out and isotope selective photodissociation. Therefore, these models are agnostic to what process caused the CO sequestering which could be caused by e.g., chemical conversion or locking of CO ice in the disk midplane. The cavity size is fixed to 60~au, the largest cavity size included in the modelling grid and close to the observed cavity size in the PDS~70 disk where the continuum emission peaks at 74~au and shows a shoulder of emission at $\sim60$~au \citep{Keppler2019, Benisty2021}. 

The priors for the disk mass, peak CO abundance, cosmic ray ionization rate, and the gas-to-dust mass ratio are flat in log-space while that for the stellar luminosity is a Gaussian centered at the measured stellar luminosity, with a width corresponding to the uncertainty on that measurement, and limited between 0.09 and 1.1~L$_{\odot}$. The stellar luminosity is included in the fitting to interpolate between the stellar spectra that are included in the DALI grid but do not match that of PDS~70 exactly. This fitting is done by linearly interpolating between the stellar luminosity, effective temperature, mass accretion rate, and stellar mass for the different stellar spectra, 'pale', 'faint', 'intermediate', and 'bright', that are included in the underlying modelling grid (see Table~1 in \citealt{Trapman2025_agepro}). Similarly, the priors for the vertical structure and dust populations are uniform in linear space and linearly interpolate between the 'flat' and 'flared' vertical structures and the 'young' and 'evolved' dust populations. Finally, the prior for $R_{\mathrm{c}}$ is uniform in linear space. The disk mass is allowed to vary between $10^{-4}-0.5$~M$_{\odot}$ following \citet{Trapman2025_exoALMA} and the prior for characteristic radius is between 15 and 300~au. Both of these allow for a slight extrapolation compared to the underlying modelling grid that contains disk masses of $10^{-4}-10^{-1}$~M$_{\odot}$ and $R_{\rm c} = 30-180$~au. This extrapolation does not substantially change our results for the fiducial case. The other parameters use the full extent of the grid as defined in \citet{Trapman2025_exoALMA, Trapman2025_agepro}. 
The posterior distribution is sampled with the \texttt{mcmc} method implemented in the \texttt{emcee} package \citep{emcee}. The \texttt{mcmc} is run with 50 walkers for  3000 steps where the first 600 steps are used for the burn-in phase (see App.~\ref{app:mcmc} for details). Finally, the chains are thinned by a factor of 100.

\begin{table}
\caption{Line constants used in this work. \label{tab:cdms}}
\centering
\begin{tabular}{lccccc}
\hline\hline
Molecule & Transition & Freq. & $\log_{10}A_{ul}$ & $g_u$ & $E_u$ \\
 &  &  (GHz) & ($\log_{10}$ s$^{-1}$) & & (K) \\
\hline
\ce{N2H+} & $1-0$ &  93.173 & $-4.4$ & 27 &  4.5  \\
\ce{N2H+} & $3-2$ & 279.512 & $-2.9$ & 63 & 26.8  \\
\hline
\end{tabular}
\tablefoot{Line constants are taken from the CDMS database \citep{Muller2001, Muller2005,Cazzoli2012, Endres2016}. }
\end{table}

As there is a wealth of ALMA data available for the PDS~70 disk, we run the \texttt{mcmc} using various combinations of the disk integrated fluxes for \ce{^12CO} $J=2-1$ and $J=3-2$, \ce{^13CO} $J=2-1$, \ce{C^18O} $J=2-1$, \ce{C^17O} $J=3-2$, and \ce{N2H+} $J=3-2$, and the 231, 273, and 337~GHz continuum emission, where the latter four are not observed directly but scaled from observations at a different frequency. In addition, we test the effect of including the 90\% flux radius of the \ce{^13CO} $J=2-1$ emission measured by \citet{Law2024} in the fitting procedure. By default, all fluxes in Table~\ref{tab:fit_inputs} and the 90\% flux radius of $167\pm 5$~au for the \ce{^13CO} $J=2-1$ emission \citep{Law2024} are used in the fitting but the effect of only using a subset is further discussed in Sect.~\ref{sec:fluxes_to_fit}.

As the continuum observations in the PDS~70 disk are at somewhat different frequencies than those included in the model grid, the observed 225, 253, and 351~GHz continuum fluxes are scaled to a frequency of 231, 273, and 337~GHz, respectively. A spectral index of $2.8\pm0.4$ is used, which is appropriate for the outer disk observed with ALMA Band~6 and 7 \citep{Doi2024, Fasano2025, Sierra2025}.

Similarly, the \ce{N2H+} $J=1-0$ line covered in the ALMA Band~3 data is not included in the DALI model grid that only includes the $J=3-2$ and $J=4-3$ transitions. Therefore, the line flux of the $J=1-0$ transition is scaled to that of the $J=3-2$ transition assuming local thermodynamical equilibrium, optically thin emission, and using the line constants from the CDMS database that does not separate between the hyperfine components \citep{Goldsmith1999, Muller2001, Muller2005, Cazzoli2012, Endres2016}:
\begin{align}
    F\Delta V_{3-2} = F\Delta V_{1-0} \frac{g_3A_{3-2}}{g_1A_{1-0}} e^{(E_1-E_3)/T}, \label{eq:scaling_N2H+}
\end{align}
with $F\Delta V$ the integrated flux, $A_{ul}$ the Einstein-$A$ coefficient, $g_u$ the degeneracy of the upper energy level, and $E_u$ the upper level energy. The subscripts $u$ and $l$ refer to the upper and lower level of the transition, respectively. A summary of the line coefficients used is presented in Table~\ref{tab:cdms}. The assumed excitation temperature is 21~K based on that of the \ce{N2H+} emission in the TW~Hya disk and on chemical modelling showing that \ce{N2H+} is abundant between the CO and \ce{N2} snowlines \citep[e.g.,][]{vantHoff2017, Schwarz2019}. For this temperature, the resulting \ce{N2H+} flux is $2.12 \pm 0.4$~Jy~km~s$^{-1}$. As the derived peak CO abundance is sensitive to the assumed excitation temperature, this is briefly discussed in Appendix~\ref{app:Tex}.

\begin{figure}
\centering
\includegraphics[width=\hsize]{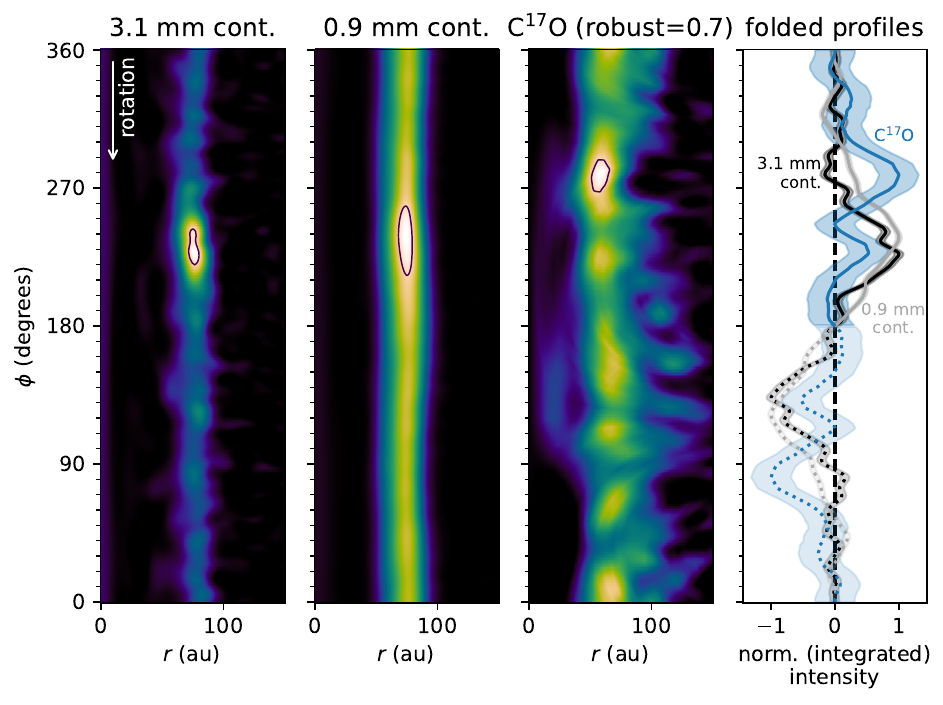}
  \caption{Polar deprojection of the emission in the PDS~70 disk starting from in the minor axis on the north east (far) side of the disk in anticlockwise direction. The three left most panels present the intensity of the 3.1~mm and 0.9~mm continuum and integrated intensity of the \ce{C^17O} $J=3-2$ line emission with a circular beam in the frame of the disk, respectively \citep[this work]{Isella2019}. The black contour is set to 90\% of the peak emission in each panel. The fourth panel shows the corresponding folded and normalized azimuthal profiles extracted from the radial location where the emission peaks (77~au for the 3.1~mm continuum in black; 74~au for the 0.9~mm continuum in grey, and 56~au for the \ce{C^17O} emission in blue). }
     \label{fig:polar}
\end{figure}

\begin{figure*}
\centering
\includegraphics[width=\hsize]{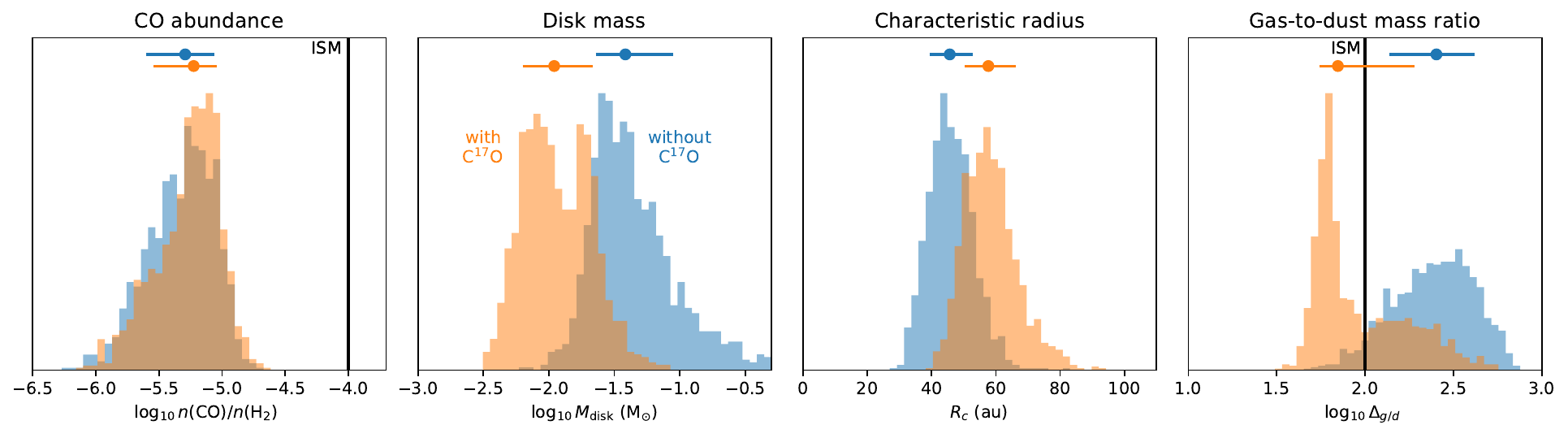}
  \caption{Posterior distribution of the peak CO abundance, disk mass, characteristic radius, and gas-to-dust mass ratio in the PDS~70 disk when fitting for all fluxes in Table~\ref{tab:fit_inputs} and $R_{\rm 90\%}$(\ce{^13CO}) (orange) or only excluding the \ce{C^17O} flux from the fitting (blue). The errorbars at the top of the figure indicate the median value of the distribution with the 16$^{\rm th}$ and 84$^{\rm th}$ percentiles as the uncertainty. }
     \label{fig:mcmc}
\end{figure*}

\section{Results} \label{sec:results}
\subsection{First detection of \ce{N2H+} and \ce{C^17O} in the PDS~70 disk}

The integrated intensity maps of the \ce{N2H+} and \ce{C^17O} emission in the PDS~70 disk are presented in Fig.~\ref{fig:mom0_N2H+} together with the 3.1~mm continuum emission. The moderate angular resolution of the \ce{N2H+} emission precludes a detailed morphological analysis but the emission appears to originate from a region with a similar radial extent as the 3.1~mm continuum and the \ce{C^17O} emission. The \ce{C^17O} emission has a higher spatial resolution than the \ce{N2H+} observations and is ring shaped with a clear central hole, similar to that seen in e.g., \ce{^13CO} and \ce{C^18O} \citep[see Fig.~\ref{fig:azi_avg_all}, ][]{Facchini2021, Rampinelli2024}. The intensity weighted velocity maps (moment~1 map) in Fig.~\ref{fig:mom1_N2H+} show that the \ce{C^17O} line emission follows Keplerian rotation, further demonstrating that the emission is associated with the PDS~70 disk. The moment~1 map of the \ce{N2H+} emission cannot be interpreted as a velocity field due to the wide separation of the hyperfine components that are not spectrally resolved at the resolution of the data. 

In addition to the central cavity, the 3.1~mm continuum emission shows a pronounced overbrightness in the northwest of the disk and a similar feature is tentatively detected in the \ce{C^17O} emission (Fig.~\ref{fig:mom0_N2H+}). This is also seen in the polar deprojection of the 3.1~mm and 0.9~mm continuum intensity and the \ce{C^17O} integrated intensity map that are presented in Fig.~\ref{fig:polar}. All three images have a beam that is circular in the frame of the disk to ensure a uniform resolution element in the polar plot. The 0.9~mm continuum image from \citet{Isella2019} and the 3.1~mm continuum image were smoothed to a $0\farcs14 \times 0\farcs087\ (-19.6\degree)$ beam using the \texttt{casa} task \texttt{imsmooth}. As the position angle of the \ce{C^17O} beam is nearly perpendicular to the beam shape that is circular in the frame of the disk, the line was reimaged with a smaller robust parameter of 0.7 and a $0\farcs179 \times 0\farcs05\ (-16.0\degree)$ $uv$-taper, resulting in a $0\farcs23 \times 0\farcs14\ (-18.6\degree)$ beam. The rightmost panel of Fig.~\ref{fig:polar} displays the folded profiles extracted at the radius of peak emission in the three left panels. The folded profiles are calculated by taking the azimuthal profile of the (integrated) intensity and subtracting the corresponding profile obtained from the disk mirrored in its minor axis. This improves the localization of the emission peak. As the beam is circular in the frame of the disk and because the central cavity is resolved, the uncertainty on the folded profiles is estimated as $\sqrt2 \sigma_{\rm mom0}$, with $\sigma_{\rm mom0}$ the rms in the moment 0 map of the non-JvM corrected cube.

The azimuthal asymmetry in the continuum emission at these wavelengths was first reported by \citet{Keppler2019} and \citep{Benisty2021} in ALMA Band~7 and is particularly pronounced at longer wavelengths in ALMA Band~3 observations \citep{Doi2024}. Interestingly, the tentative asymmetry ($\sim 3\sigma$) in the folded profile of the \ce{C^17O} emission is shifted compared to that in the continuum emission. As the disk is rotating clockwise in the plane of the sky, the continuum emission peaks $\sim 40-50\degree$ ahead of the \ce{C^17O} emission in the direction of rotation. This asymmetry is not seen in the $J=2-1$ transition of the more abundant \ce{^12CO}, \ce{^13CO}, and \ce{C^18O} isotopologues presented in \citet{Rampinelli2024}, see Fig.~\ref{fig:polar_COiso}.

\subsection{peak CO abundance and disk mass }

The adopted model grid follows the framework used in previous studies \citep[e.g.,][]{Trapman2025_exoALMA, Trapman2025_agepro}, allowing us to derive disk parameters for the PDS~70 disk that can be directly compared with those inferred for other disks using the same methodology.
The results of the \texttt{mcmc} sampling are presented in Fig.~\ref{fig:mcmc}, where the posterior distributions for the peak CO abundance, disk mass, $R_{\mathrm{c}}$, and the gas-to-dust mass ratio are shown. The other parameters that are fitted for (disk flaring, dust size population, cosmic ray ionization rate, and the stellar luminosity) are not meaningfully constrained by the \texttt{mcmc} sampling procedure. 
The orange posteriors are the result of the fiducial set of line fluxes that are included in the fitting (see Table~\ref{tab:fit_inputs}) together with $R_{\rm 90\%} ($\ce{^13CO}). The blue histograms present the posterior distributions for the same set of input fluxes and $R_{\rm 90\%} ($\ce{^13CO}) except that the \ce{C^17O} line flux is not included in the fitting.

Figure~\ref{fig:mcmc} demonstrates that the \texttt{mcmc} sampling procedure prefers a different family of models depending on whether \ce{C^17O} is included in the fit. 
The fit including the \ce{C^17O} line flux prefers a more extended disk with $R_c = 58^{+9}_{-7}$~au and a disk mass of $M_{\rm disk} = 0.011^{+0.011}_{-0.005}$~M$_{\odot}$, where the uncertainties are the 16$^{\rm th}$ and 84$^{\rm th}$ percentiles of the posterior distribution and do not include any systematic uncertainties. In contrast, the fit excluding the \ce{C^17O} emission favors a smaller disk with $R_c = 46^{+7}_{-6}$~au and disk mass that is $\sim 3.4\times$ lower. The gas-to-dust mass ratio shifts by roughly the same factor as the disk mass but in opposite direction when the \ce{C^17O} flux is included. Regardless of this shift, the tail of the $\Delta_{\rm g/d}$ distribution is consistent with the ISM value. These changes in the preferred model parameters are likely driven by a reweighting of the broad and degenerate family of models favored by the \texttt{mcmc} sampling, as the predicted \ce{C^17O} flux is close to the observed value for the \texttt{mcmc} chains \textit{excluding} that line in the fitting (see App.~\ref{sec:mcmc_fluxes}).

In contrast to the disk mass, gas-to-dust mass ratio, and $R_{\mathrm{c}}$, the peak CO abundance is more or less independent of whether the \ce{C^17O} line flux is included in the fitting. This is because the CO depletion is mainly set by the \ce{N2H+} flux. We derive a peak CO abundance of $(5.1^{+3.6}_{-2.6} - 6.0^{+3.2}_{-3.1}) \times 10^{-6}$ \ce{H2} assuming that the \ce{N2H+} emission is optically thin and is characterized by an excitation temperature of 21~K. This value changes by less than a factor of $\sim 3$ if the \ce{N2H+} is characterized by a temperature of 11-31~K or if the emission is optically thick (see App.~\ref{app:Tex}). In summary, this indicates that CO is sequestered over approximately an order of magnitude compared to the ISM abundance after accounting for CO freeze-out and isotope selective photodissociation.

However, the inferred disk parameters are conditional on the assumptions of the adopted model grid, which does not account for the gas cavity observed in the PDS~70 disk. This structural mismatch may introduce potentially substantial systematic uncertainties that are not captured by the posterior distributions. We discuss the implications of this limitation in Sect.~\ref{sec:discussion_systematics}.

\begin{figure}
\centering
\includegraphics[width=\hsize]{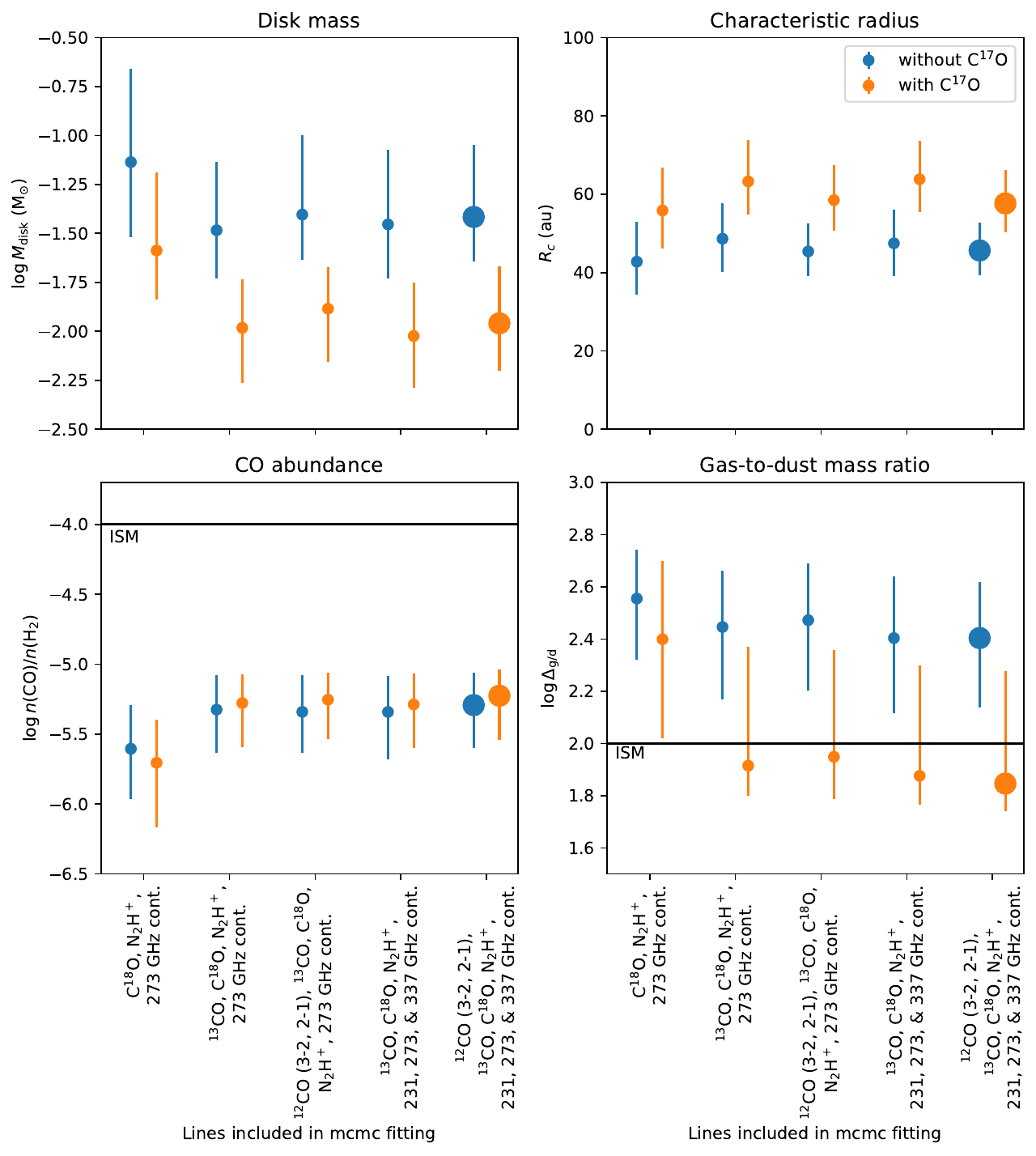}
  \caption{Derived peak CO abundance, disk mass, gas-to-dust mass ratio, and $R_{\mathrm{c}}$ for various combinations of lines fluxes that are fitted for in the \texttt{mcmc} sampling. The orange datapoints all include \ce{C^17O} in the fit whereas the blue datapoints present the results of the \texttt{mcmc} sampling without the \ce{C^17O} line flux. The two scatterpoints with a larger markersize indicate the fiducial case with and without \ce{C^17O}. }
     \label{fig:fluxes_to_fit}
\end{figure}

\begin{figure*}
\centering
\includegraphics[width=\hsize]{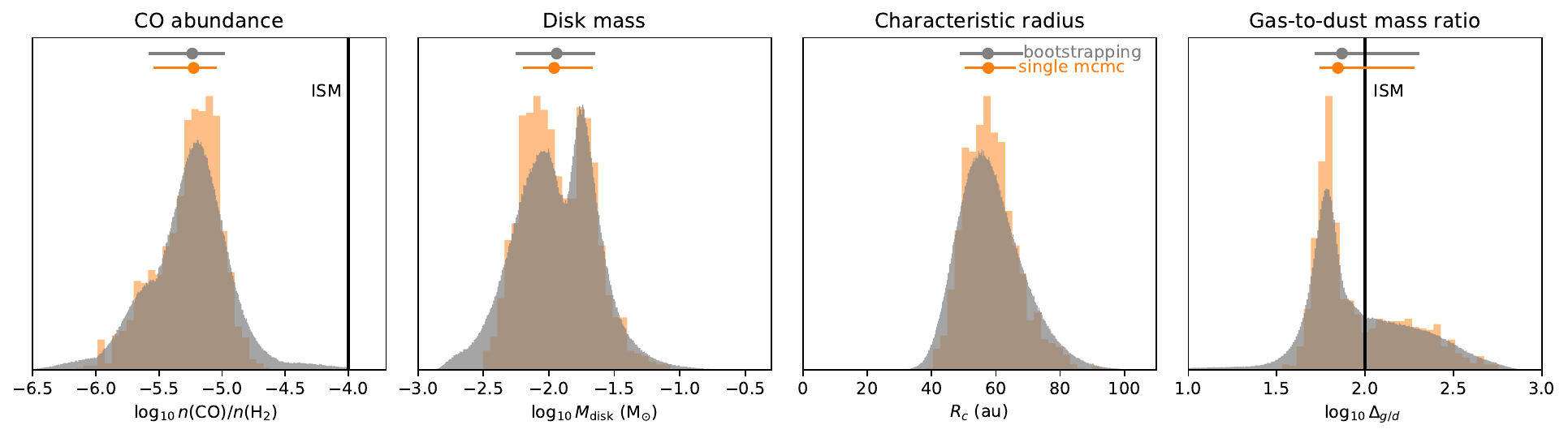}
  \caption{Distribution of the posteriors for derived peak CO abundance, disk mass, $R_{\mathrm{c}}$, and gas-to-dust mass ratio for 1000 \texttt{mcmc} runs with input fluxes (including the \ce{C^17O} flux) and $R_{\rm 90\%}$(\ce{^13CO}) values sampled from the measured fluxes (grey). The orange histograms are identical to those in Fig.~\ref{fig:mcmc} and show the result of a single \texttt{mcmc} fit using the measured values and their associated uncertainties.}
     \label{fig:bootstrapping}
\end{figure*}

\subsection{Systematics in $M_{\rm disk}$, $R_{\rm c}$, and $\Delta_{\rm g/d}$} \label{sec:fluxes_to_fit}

The systematic shift in the disk mass, $R_{\mathrm{c}}$, and gas-to-dust mass ratio when including \ce{C^17O} in the \texttt{mcmc} sampling indicates that the derived disk parameters could depend on which fluxes are included in the fitting. For example, optically thick lines primarily trace the temperature and emitting area, whereas optically thin lines trace the temperature and the total column of CO molecules. Rarer CO isotopologues will probe deeper and thus colder layers in the disk. To further investigate the effect of which fluxes are included in the fitting, we ran the fitting for various combinations of CO isotopologues and continuum fluxes but always including the \ce{C^18O} and \ce{N2H+} line flux. In addition, for each run, the effect of the \ce{C^17O} line flux is investigated. The results are presented in Fig.~\ref{fig:fluxes_to_fit} where the runs including the \ce{C^17O} line flux are shown in orange whereas those without the \ce{C^17O} are shown in blue. 

This figure demonstrates that the disk mass and gas-to-dust mass ratio are systematically lower if \ce{C^17O} is included in the \texttt{mcmc} sampling regardless of the other fluxes included in the fitting. The only exception to this are the runs fitting just the \ce{C^18O}, \ce{N2H+}, and the 273~GHz continuum with and without the \ce{C^17O} line flux. The systematic shift indicates that the observed \ce{C^17O} line flux is lower than expected from the models that fit to the other (line) fluxes. A possible cause for that could be that the CO isotopologues in the PDS~70 disk emit from a colder layer than predicted by the models, lowering the flux of primarily the optically thick CO isotopologues. Another possible cause is the lack of a gas gap in the models. This is further discussed in Sect.~\ref{sec:discussion_systematics}. Only the derived peak CO abundance is independent of the fluxes used in the fitting as this is mainly driven by the \ce{N2H+} flux. Fitting for $R_{\rm 90\%}$(\ce{^13CO}) primarily constrains $R_c$ but does not have a large effect on the derived disk mass, peak CO abundance, and gas-to-dust mass ratio as discussed in Appendix~\ref{app:R90}.

\subsection{Bootstrapping} \label{sec:bootstrapping}

The systematic shift in the derived disk mass depending on the inclusion of the \ce{C^17O} line flux suggests that the uncertainties of the \texttt{mcmc} sampling procedure underestimate the true uncertainty on the derived parameters. To investigate this, 1000 random samples of CO isotopologue, \ce{N2H+}, and continuum fluxes and of $R_{\rm 90\%}$(\ce{^13CO}) are drawn from a Gaussian probability distribution representing the measured values and their associated uncertainties following e.g., \citet{Longarini2025} and \citet{Pezzotta2026}. These values are then used in the \texttt{mcmc} sampling procedure to better assess the uncertainty on the derived parameters, such as, the disk mass and peak CO abundance due to the uncertainty on the measured line fluxes and radius. 

The resulting distribution of all 1000 posteriors is presented in Fig.~\ref{fig:bootstrapping} where the \ce{C^17O} flux is included in the fit (see Fig.~\ref{fig:bootstrapping_noC17O} for the figure without fitting for the \ce{C^17O} flux). This distribution is similar to that of the posterior distribution of the single \texttt{mcmc} run using the measured values and their uncertainties. This indicates that the uncertainty on the measured line and continuum fluxes and the 90\% flux radius of the \ce{^13CO} emission are not the main source of uncertainty on the derived parameters such as the disk mass. Instead, the underlying modelling grid is a main source of the spread in the derived values. All in all, we derive a disk mass of $0.011^{+0.011}_{-0.006} - 0.039^{+0.067}_{-0.018}$~M$_{\odot}$ where the range is due to the \ce{C^17O} flux and assuming that the \ce{N2H+} is optically thin and is characterized by an excitation temperature of 21~K. The peak CO abundance in this case is $(4.8^{+4.3}_{-2.7} - 5.8^{+4.7}_{-3.2}) \times 10^{-6}$ with respect to \ce{H2}. If the \ce{N2H+} emission is optically thick instead, the peak CO abundance can reach up to $(1.4^{+1.0}_{-0.7} - 1.7^{+0.8}_{-0.8}) \times 10^{-5}$ with respect to \ce{H2}. The derived peak CO abundance is not necessarily the global CO/\ce{H2} ratio in the disk but rather the peak CO abundance in the emitting layer probed by the observations. In addition, the value itself depends on the assumed scaling between $1-0$ and $3-2$ transition of \ce{N2H+} (see App.~\ref{app:Tex}), the ionization prescription, and the chemical network. We note that the latter includes CO photodissociation and freeze-out but no grain surface chemistry nor locking of ices in the disk midplane.

\section{Discussion} \label{sec:discussion}

\subsection{Origin of the asymmetry in the \ce{C^17O} emission}

The marginal azimuthal asymmetry in the \ce{C^17O} emission that is lagging behind that seen in the (sub-)mm continuum emission in the rotation direction of the disk could hint at a vortex at that location. Theoretical work modelling the interaction between the gas and dust in a vortex generated by the Rossby Wave Instability shows that the peak of the dust density shifts ahead of the gas overdensity \citep{Lovelace1999, Mittal2015, Baruteau2016, Ono2016}. In particular, when self-gravity is included, grains with large Stokes numbers are predicted to show a larger shift up to 90$\degree$ than grains with smaller Stokes numbers. This is qualitatively similar to the trends seen in the PDS~70 data as the \ce{C^17O} is the rarest and thus most optically thin CO isotopologue observed in this disk and thus is the closest tracer of the gas surface density. However, new deep observations of even more rare CO isotopologues and/or a detailed analysis of the kinematics at high spatial resolution in this region are necessary to confirm if the \ce{C^17O} emission is indeed tracing a vortex. 

Azimuthal asymmetries possibly related to vortices have been found in a number of other disks. Multi-wavelength analyses have shown that the spectral index drops in sub-mm crescents compared to other dusty disk regions indicating an increased dust density in the asymmetry (e.g., the PDS~70, HD~34282, HD~135344B, and HD~142527 disks; \citealt{Doi2024, Ma2026, Cazzoletti2018, Soon2019}, respectively). Even though the origin of these asymmetries is unclear in general, the similarity in azimuthal velocity between dust asymmetry and the PDS~70c planet suggests that this planet could be triggering the dust asymmetry \citep{Fasano2025}. In addition to evidence for radial and azimuthal trapping of dust through continuum observations, vortices could also leave an imprint in the gas kinematics \citep[e.g.,][]{Robert2020, Wolfer2025}. Despite tentative evidence for kinematical signatures in some disks, unambiguous evidence is still missing \citep[e.g.,][]{Yen2020, Boehler2021, Wolfer2025, Stadler2026}. Similarly, no signatures of rotating gas influencing the temperature structure were found around the mm-crescent in the IRS~48 disk \citep{Temmink2025}. In summary, despite multiple disks showing azimuthal asymmetries possibly caused by vortices, conclusive and unambiguous evidence is still lacking.

\subsection{PDS~70 disk mass} \label{sec:discussion_Mdisk}

The \texttt{mcmc} sampling shows that the fluxes in the PDS~70 disk are consistent with models with a disk mass of $0.011^{+0.011}_{-0.006} - 0.039^{+0.067}_{-0.018}$~M$_{\odot}$ ($12^{+12}_{-6} - 41^{+70}_{-19}$~M$_{\rm Jup}$ or $M_{\rm disk}/M_{\star} = 1.3^{+1.3}_{-0.7} - 4.5^{+7.7}_{-2.1}$~\%) and a gas-to-dust mass ratio of $\sim 74^{+128}_{-22} - 256^{+186}_{-129}$ if the \ce{N2H+} emission is optically thin and characterized by an excitation tempeature of 21~K. As the shift in the gas-to-dust mass ratio acts in the opposite direction to the shift in disk mass, both families of models predict that the PDS~70 dust mass is $\sim 50$~M$_{\oplus}$. 
This is consistent with the lower limit of $>28$~M$_{\oplus}$ on the dust mass derived from ALMA Band~3 observations and is consistent with the lower end of the dust mass estimates from multi-wavelength analysis that indicate a dust mass of $\sim 10^{-4}-10^{-3}$~M$_{\odot}$ depending on the grain size distribution \citep{Benisty2021, PortillaRevelo2022, Doi2024, Sierra2025}. A detailed study combining a multi-wavelength analysis with the observed polarization in this disk indicates a somewhat higher dust mass of 281~M$_{\rm Earth}$ \citep{Liu2026}. 

To our knowledge only two estimates of the disk mass based on thermochemical modelling are published. \citet{Cridland2023} modelled the radial profiles of the \ce{^12CO} and \ce{C^18O} $J=2-1$ lines and the \ce{C2H} emission and find a disk mass of $1.5\times 10^{-3}$~M$_{\odot}$, which is an order of magnitude lower than the range of masses that we find. A possible cause for this could be the low dust mass of 5~M$_{\oplus}$ or the ISM like C/H in their model rather than modelling the CO sequestering by lowering the elemental C/H abundance. Our range of disk mass estimates is consistent with the lower limit of $>3\times 10^{-4}$~M$_{\odot}$ derived by \citep{PortillaRevelo2023}. 

\subsection{Systematic uncertainties in the disk mass} \label{sec:discussion_systematics}

The family of models preferred by the \texttt{mcmc} sampling procedure depends on whether the \ce{C^17O} line flux is included in the fitting, independent of which other fluxes are used in the fit (see Sect.~\ref{sec:fluxes_to_fit}). The run including the \ce{C^17O} line emission prefers a lower mass disk that has a somewhat larger $R_{\mathrm{c}}$ than if the \ce{C^17O} is not included. The inclusion of the \ce{C^17O} flux therefore changes the relative weights of the models within the broad and degenerate posterior distribution, even though the predicted \ce{C^17O} fluxes are close to the observed value when this line is excluded from the fitting (see App.~\ref{sec:mcmc_fluxes}). Two possible reasons for this include a difference between the temperature structure of the PDS~70 disk and the models and the central gas gap that is not present in the models. The former primarily affects the line flux of the optically thick CO isotopologues as those trace a combination of temperature and area, whereas the optically thin CO isotopologues primarily trace the column density rather than the temperature.

The second source of uncertainty is the gas gap that is seen in the PDS~70 disk but not included in the models that only have a dust cavity. High resolution observations of the \ce{^12CO} $J=3-2$ transition that spatially resolve the dust cavity show that there is a gap rather than a full gas cavity in the PDS~70 disk \citep[see Fig.~\ref{fig:azi_avg_all};][]{Keppler2019, Facchini2021}. Similarly, the \ce{HCO+} $J=4-3$ transition also shows some emission in the inner disk \citep{Long2018, Facchini2021}. This is further supported by \citet{PortillaRevelo2023} who find that the gas density drops by a factor of $\sim 19$ in the gap using thermochemical models. 

An upper limit on the mass carried in the gas gap can be estimated by integrating the radial gas surface density profile used in the \texttt{DALI} thermochemical models for best fitting parameters. The mass carried inside 45~au, the location where the \ce{^12CO} $J=3-2$ transition peaks outside the gas gap, is $< 7 - 26$~M$_{\rm Jup}$ or $< 54-63$\% of the total disk mass. Even though this is roughly consistent with the combined estimated masses of PDS~70b and c \citep[e.g.,][]{Keppler2018, Haffert2019, Wang2020, Doi2024, Trevascus2025}, we emphasize that this is not a measurement of the gas mass inside the gas cavity in the PDS~70 disk but rather should be considered as an additional source of uncertainty on the derived disk mass as the modelling grid does not include a gas cavity. 

To further investigate the sensitivity of the derived disk parameters to the \ce{C^17O} flux, we performed a sensitivity test where an additional \texttt{mcmc} sampling of the fiducial case was run with the \ce{C^17O} flux increased by a factor of 2. This choice is motivated by the difference in optical depth between the CO isotopologues. \ce{C^17O} is the most optically thin tracer and thus most affected by the mass whereas the other (more) optically thick lines of e.g., \ce{^12CO} and \ce{^13CO} mainly trace the surface area that only changes by a small amount as the gas gap is much smaller than the total area of the disk. The result of this sensitivity test is presented in Fig.~\ref{fig:mcmc_C17Ox2}. Interestingly, the sampling with the doubled \ce{C^17O} flux almost exactly matches the posteriors when \ce{C^17O} is not included in the fitting, highlighting that the derived disk mass, size and gas-to-dust mass ratio are sensitive to a factor of 2 change in the \ce{C^17O} flux that is only one out of nine fluxes that are fitted for. Despite this close match, this sensitivity test does not provide a basis for favoring the \texttt{mcmc} sampling without \ce{C^17O} over that including the \ce{C^17O}. Importantly, doubling the integrated \ce{C^17O} flux does not physically reproduce the effects of a gas gap, which would also alter the temperature, shielding against UV radiation of the star, molecular excitation, and spatial distribution of the molecular emission. Therefore, this experiment illustrates the sensitivity of the inferred parameters to the \ce{C^17O} flux rather than providing a quantitative estimate of the systematic uncertainty associated with the missing gas gap. A model grid with a more flexible (gas) surface density profile needs to be considered to get an accurate measurement of the disk mass, especially for disks with large central gaps.

Although the adopted model grid allows us to compare the inferred gas mass and CO abundance of the PDS~70 disk with those derived for other disks using the same methodology, the gas gap observed in the PDS~70 disk highlights an important limitation of the assumed disk structure. The resulting systematic uncertainties are not captured by the formal posterior distributions and may also affect gas-mass estimates for other disks, particularly when the observations do not spatially resolve gas gaps or cavities. A more comprehensive model grid that includes a range of gas gap sizes and depletion factors is therefore needed to quantify these systematic effects and improve the accuracy of disk gas-mass estimates.

\begin{figure}
\centering
\includegraphics[width=\hsize]{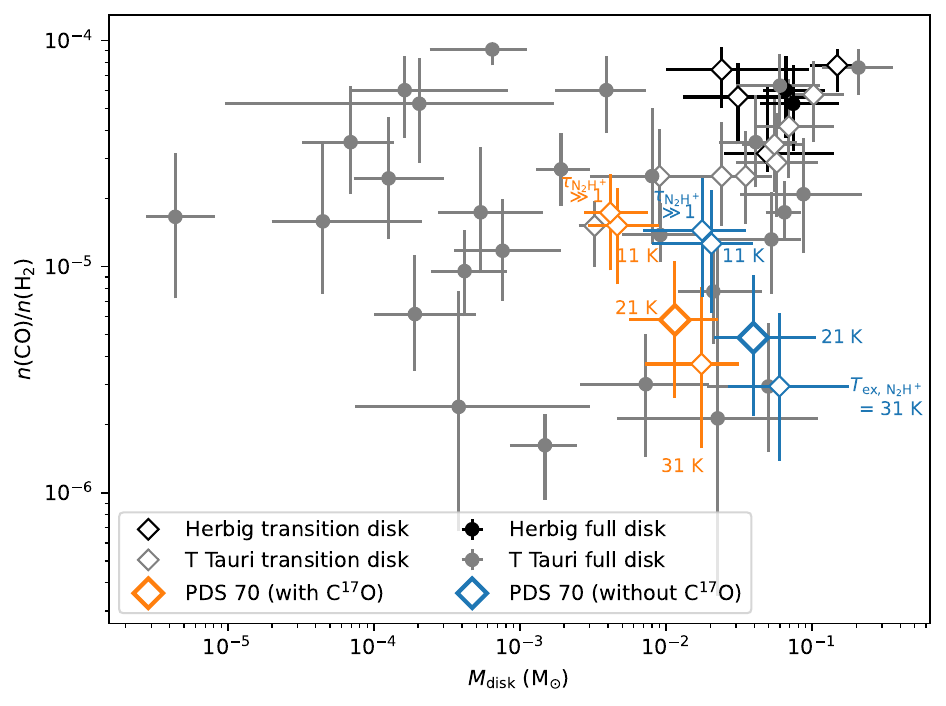}
  \caption{peak CO abundance and disk mass of all disks analyzed with the framework presented in \citet{Trapman2022, Trapman2025_exoALMA, Trapman2025_agepro}. The different datapoints for the PDS~70 disk show the derived peak CO abundance and gas mass with and without \ce{C^17O} in the \texttt{mcmc} fitting and for various scalings of the \ce{N2H+} $1-0$ to $3-2$ line flux. For TW~Hya, only a range was given in \citet{Trapman2022}. }
     \label{fig:other_disks}
\end{figure}

\subsection{Comparison to other disks}
An overview of the peak CO abundance and disk mass of all disks analyzed using the framework presented in \citet{Trapman2022, Trapman2025_exoALMA, Trapman2025_agepro} is presented in Fig.~\ref{fig:other_disks}. In this figure, transition disks are defined as all disks with a resolved (sub-)mm cavity $> 10$~au and full disks have unresolved cavities in the (sub-)mm or dust cavities $<10$~au. In addition to these two works, this figure also includes the analysis of the exoALMA program and on the young GY~91 disk \citep{Trapman2025_exoALMA, Jiang2026}. The inferred disk mass and peak CO abundance of the PDS~70 disk are broadly consistent with the values derived for the T~Tauri disk population analyzed using the same methodology, irrespective of the assumptions regarding the optical depth and excitation temperature of the \ce{N2H+} emission or the inclusion of the \ce{C^17O} line flux in the \texttt{mcmc} fitting. Other systematic uncertainties on the effective peak CO abundance in the emitting layer such as those arising from the ionization prescription and the adopted chemical network are expected to be similar between sources, thus affecting the absolute rather than relative values. In particular, the PDS~70 disk mass is similar to that of Herbig disks and massive T~Tauri disks \citep[and Fig.~\ref{fig:other_disks}]{Stapper2024}. Interestingly, if the \ce{N2H+} emission is optically thick or has a low excitation temperature of only 11~K, the PDS~70 disk is similar to other disks with large (sub-)mm cavities (diamonds scatter points). However, observations of a second \ce{N2H+} transition at sufficiently high spectral resolution to resolve the hyperfine components are needed to constrain the optical depth and excitation temperature of the \ce{N2H+} emission, thereby further narrowing the currently allowed effective peak CO abundance of $(4.8^{+4.3}_{-2.7} - 17^{+8}_{-8}) \times 10^{-6}$ relative to \ce{H2} in the layers probed by the PDS~70 disk observations and under the assumption of the adopted chemical network and ionization prescription.

\subsection{CO sequestering at the time of planet formation}

Planets accrete gas and dust with icy mantles from the protoplanetary disk they are born in. Therefore, the chemical composition of both the gas and the ice could affect the chemical composition of planetary atmospheres once planets are fully formed. In the case of the PDS~70 disk, the planets feed from the gas and ice that is delivered through the gap. The gas outside the gap is volatile poor, as traced by the CO sequestering. In addition, the outer disk gas has a high C/O ratio of $\gtrsim 1$ indicated by the detection of carbon-bearing molecules such as \ce{C2H}, \ce{c-C3H2}, and CS and the non-detection of oxygen-bearing species such as SO \citep{Facchini2021, Cridland2023, Rampinelli2024}. Using chemical evolution models of the PDS~70 system, \citet{Cridland2023} showed that the atmospheric C/O ratios of PDS~70b and c are expected to roughly reflect the composition of the disk gas from which they accreted, provided that chemical evolution occurs on timescales longer than planet formation. Meanwhile, the inner PDS~70 disk shows emission of \ce{H2O} lines with JWST \citep{Perotti2023}. In addition, radially resolved isotopologue observations of HCN suggest that fractionation processes vary across the disk, providing a potential future probe of the connection between disk chemistry and the material accreted by forming planets \citep{Zhang2021_Yapeng, Nomura2023, Rampinelli2025}.

Similar to the PDS~70 disk, many T~Tauri disks show signs of CO sequestering beyond freeze-out and photodissociation \citep[e.g.,][]{Schwarz2016, Miotello2017, Zhang2019, Zhang2021_coco, Sturm2023, Trapman2025_exoALMA, Trapman2025_agepro, PanequeCarreno2025}. Including grain surface chemistry involving CO ice can reconcile the observed CO isotopologue fluxes with model predictions, suggesting that at least part of the sequestered CO is locked in the form of e.g., \ce{CO2} ice and icy complex organic molecules \citep{Ruaud2022, Pascucci2023, Deng2025, Zwicky2025}. In addition, \citet{Bergin2024} demonstrated that some degree of carbon sequestering can help to reproduce the spread in the carbon abundance in giant planets. Current observations constrain the atmospheric [C/H] abundance of PDS~70b to $-0.2^{+0.8}_{-0.5}$ dex relative to the stellar value \citep{Hsu2024}. 
As PDS~70b and c may accrete the final portions of their atmospheres from the molecular layer in the outer disk where CO could be sequestered, this sequestration may need to be considered when interpreting the relationship between planetary atmospheric compositions and the chemistry of their natal environment.

\section{Conclusions} \label{sec:conclusions}

We have presented the first detection of \ce{N2H+} and \ce{C^17O} in the PDS~70 disk, both of which were serendipitously observed by ALMA. We used the disk integrated fluxes of these two lines together with previously published data of CO isotopologues and the continuum emission to constrain the peak CO abundance and mass of the PDS~70 disk by comparing these to a grid of thermochemical models by \citet{Trapman2025_exoALMA, Trapman2025_agepro}, allowing for a consistent comparison with other disks analyzed using the same framework. Our main findings are summarized below:

\begin{enumerate}
    \item The effective peak CO abundance relative to \ce{H2} in the PDS~70 disk layers probed by the observations is $(4.8^{+4.3}_{-2.7} - 17^{+8}_{-8}) \times 10^{-6}$, under the assumption of the adopted chemical network and ionization prescription. The range is due to the optical depth and excitation temperature of the \ce{N2H+} emission and to a lesser degree to whether the \ce{C^17O} flux is included in the fitting. This peak CO abundance is similar to what is found in other T~Tauri disks and is approximately an order of magnitude lower than that in the ISM after accounting for CO freeze-out and isotope selective photodissociation of CO. 
    \item The disk mass of the PDS~70 disk is $0.011^{+0.011}_{-0.006} - 0.039^{+0.067}_{-0.018}$~M$_{\odot}$ ($12^{+12}_{-6} - 41^{+70}_{-19}$~M$_{\rm Jup}$) assuming that the \ce{N2H+} emission is optically thin and emits at 21~K. 
    \item The range of a factor of $\sim3.4$ in the derived disk mass is driven by the dependence on whether the \ce{C^17O} $3-2$ line flux is included in the \texttt{mcmc} sampling procedure. The derived peak CO abundance is independent of this.
    \item As CO is sequestered in the outer disk gas probed by the observations, the PDS~70 planets may accrete the final portion of their atmospheres from this CO sequestered reservoir. This makes this system a good candidate to explore a direct link between the atmospheric composition of the planets and their natal environment.
    \item The large gas gap observed in the PDS~70 disk is not included in the adopted model grid and may introduce substantial systematic uncertainties in the inferred disk parameters that are not captured by the formal posterior distributions. This limitation may also affect gas-mass estimates for other disks analyzed using similar model grids, particularly when their gas cavities are not spatially resolved. More flexible model grids including a range of gas cavity sizes and depletion factors are therefore needed to quantify these systematic effects.
    \item We find tentative ($\sim 3\sigma$) evidence of an azimuthal asymmetry in the \ce{C^17O} emission that is trailing the asymmetry in the sub-mm continuum emission by $\sim 40-50\degree$. This is consistent with modelling predictions for a vortex \citep{Mittal2015, Baruteau2016} but more data are needed to confirm this.
\end{enumerate}

The CO sequestering in the first disk with a direct detection of two accreting planets demonstrates that the chemical composition of the planet forming material feeding onto the planets may vary during the planet formation process and needs to be considered when modelling the chemical composition of exoplanets. METIS and HARMONI on the ELT can further improve our understanding by characterizing exoplanetary atmospheres.

\begin{acknowledgements}
We thank the referee for the constructive comments.
M.L., S.F., L.R., and V.P. are funded by the European Union (ERC, UNVEIL, 101076613). Views and opinions expressed are however those of the author(s) only and do not necessarily reflect those of the European Union or the European Research Council. Neither the European Union nor the granting authority can be held responsible for them. S.F. acknowledges financial contribution from PRIN-MUR 2022YP5ACE. 
M.B. and L.M.S. have received funding from the European Research Council (ERC) under the European Union’s Horizon 2020 research and innovation programme (PROTOPLANETS, grant agreement No. 101002188). 
C.L. has been supported by the UK Science and Technology Research Council (STFC) via the consolidated grant ST/W000997/1.
This paper makes use of the following ALMA data: ADS/JAO.ALMA\#2022.1.01477.S ALMA is a partnership of ESO (representing its member states), NSF (USA) and NINS (Japan), together with NRC (Canada), NSTC and ASIAA (Taiwan), and KASI (Republic of Korea), in cooperation with the Republic of Chile. The Joint ALMA Observatory is operated by ESO, AUI/NRAO and NAOJ.
\end{acknowledgements}

%
%

\bibliographystyle{aa} 
\bibliography{refs.bib} 

\appendix

\section{Observational details}

\begin{figure*}
\centering
\includegraphics[width=0.8\hsize]{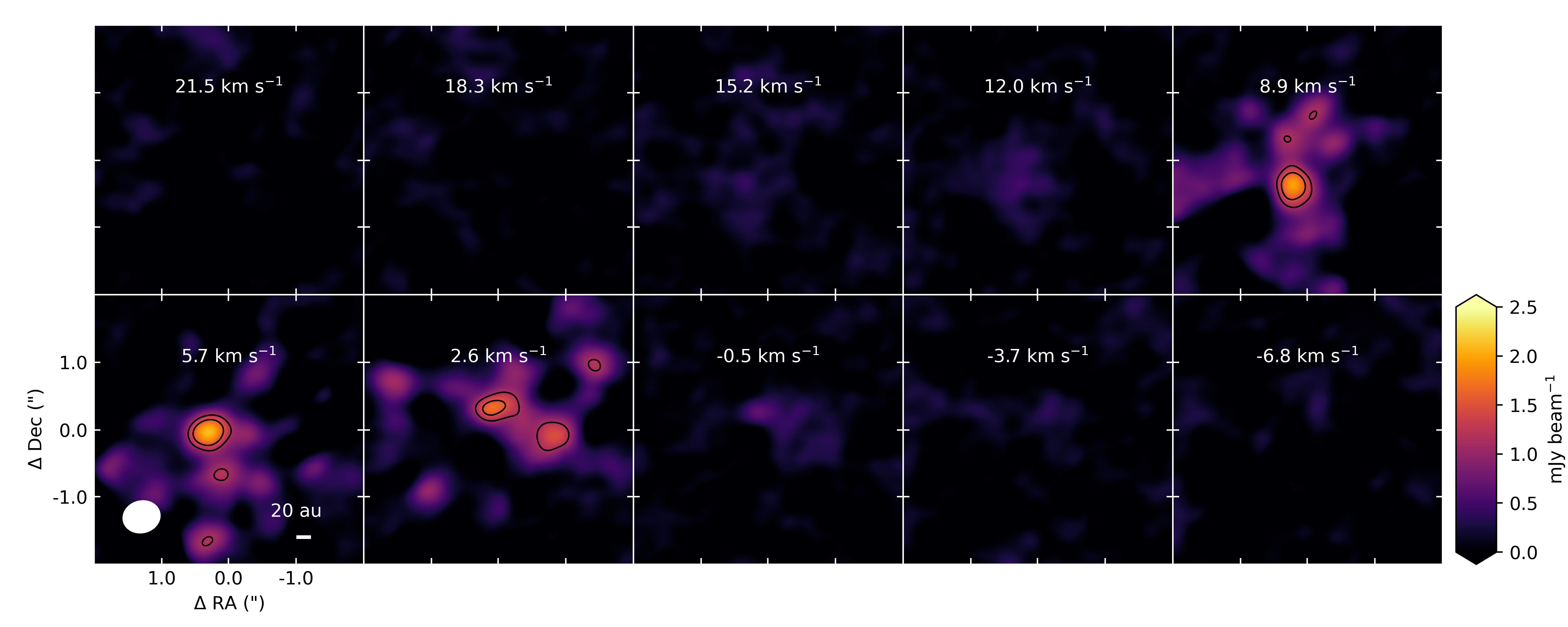}
  \caption{\ce{N2H+} channel maps. The black contours indicate the 3 and 4$\sigma$ confidence levels and the beam and a 20~au scalebar are indicated in the bottom left panel.}
     \label{fig:chans_N2H+}
\end{figure*}

\begin{figure*}
\centering
\includegraphics[width=0.8\hsize]{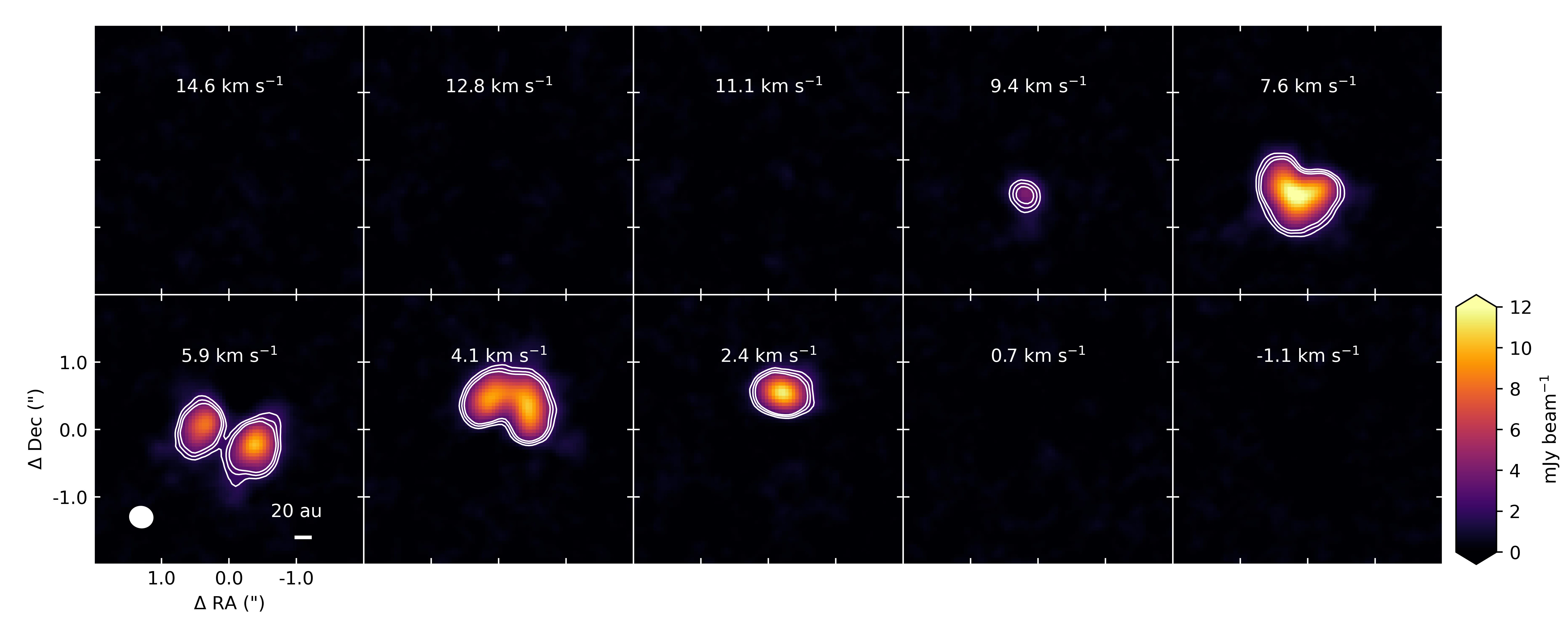}
  \caption{\ce{C^17O} channel maps. The black contours indicate the 3, 4, and 5$\sigma$ confidence levels and the beam and a 20~au scalebar are indicated in the bottom left panel.}
     \label{fig:chans_C17O}
\end{figure*}

\begin{figure}
\centering
\includegraphics[width=\hsize]{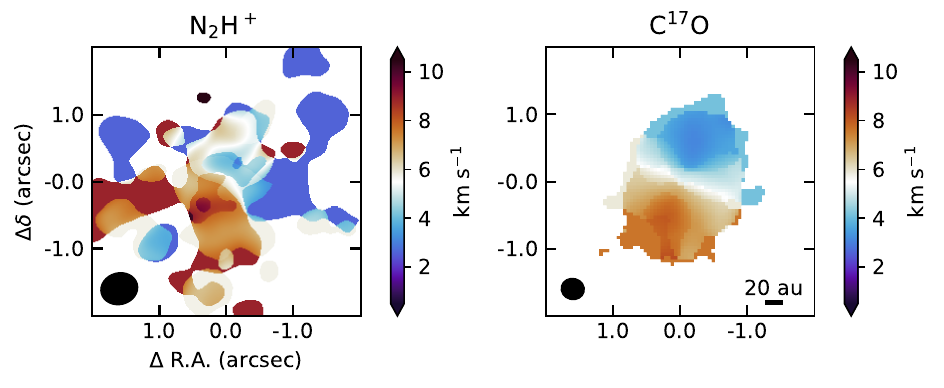}
  \caption{Intensity weighted velocity maps of the \ce{N2H+} (left) and \ce{C^17O} (right) emission in the PDS~70 disk. The beam is indicated in the bottom left corner of each panel and a 20~au scalebar is shown in the bottom right corner of the last panel. }
     \label{fig:mom1_N2H+}
\end{figure}

\begin{figure}
\centering
\includegraphics[width=\hsize]{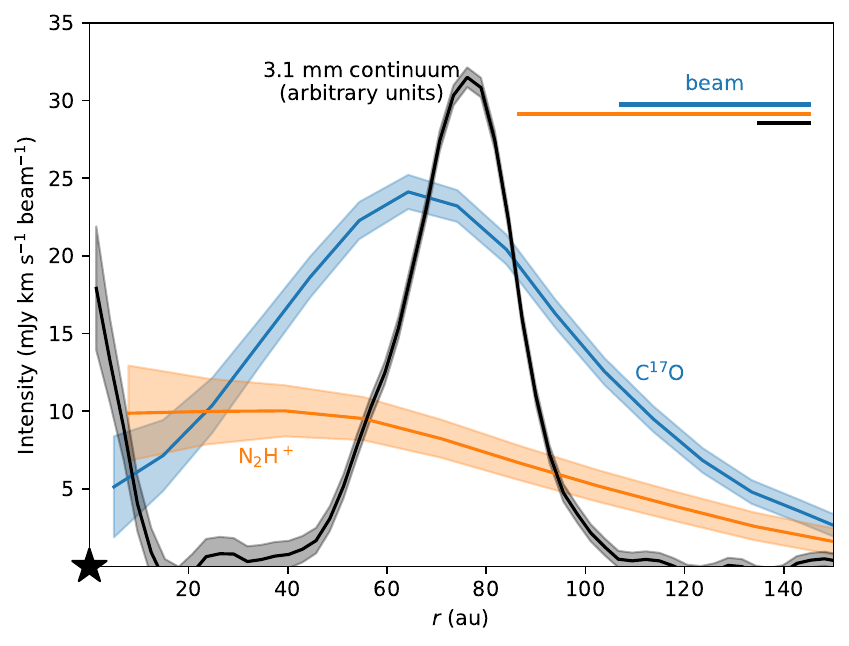}
  \caption{Azimuthally averaged radial profiles of the 3.1~mm continuum (black), \ce{N2H+} (orange), and \ce{C^17O} (blue) emission in the PDS~70 disk. }
     \label{fig:azi_avg}
\end{figure}

\begin{figure}
\centering
\includegraphics[width=\hsize]{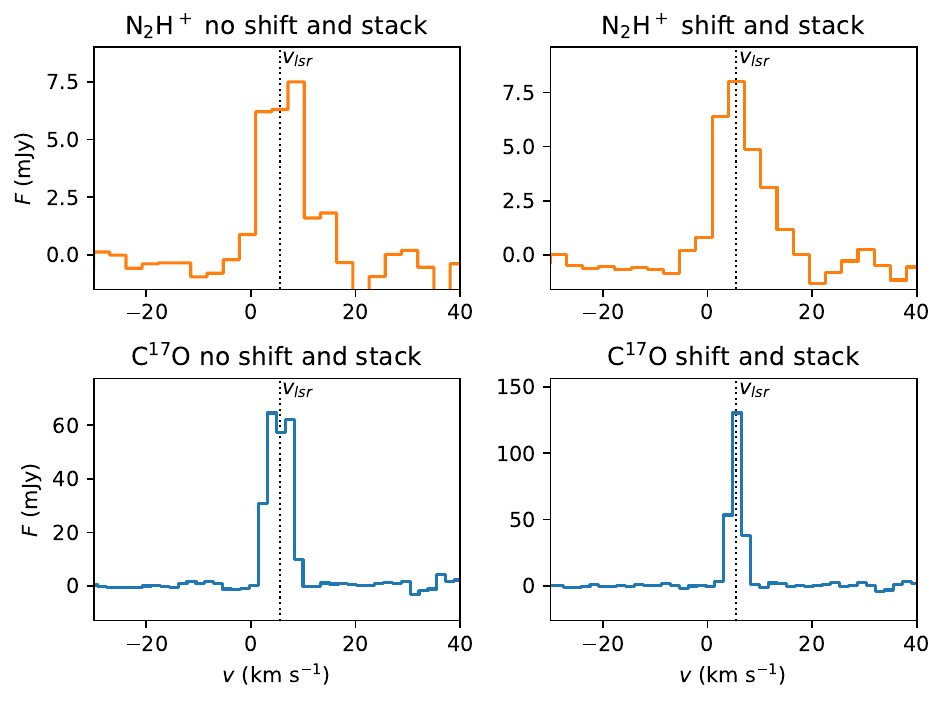}
  \caption{Spectra of the \ce{N2H+} (top) and \ce{C^17O} (bottom) line emission in a $1\farcs8\times 1\farcs1$ (semi-axes) elliptical region. The left column shows the spectrum without shifting and stacking the data whereas in the right column, the spectrum in each pixel is shifted by the projected Keplerian velocity at that location before spatially integrating the emission. The vertical black line indicates the source velocity of the PDS~70 system. }
     \label{fig:spectra}
\end{figure}

A summary of the EBs covering the \ce{N2H+} $1-0$ and the \ce{C^17O} $3-2$ transitions is presented in Table~\ref{tab:obs_details}. The channel maps of both lines are presented in Fig.~\ref{fig:chans_N2H+} and \ref{fig:chans_C17O}, respectively. The moment~1 maps can be found in Fig.~\ref{fig:mom1_N2H+} and the azimuthally averaged radial profiles are presented in Fig.~\ref{fig:azi_avg}. Finally, the spectra are shown in Fig.~\ref{fig:spectra}. The left column of Fig~\ref{fig:spectra} shows the spectrum integrated over an $1\farcs8\times 1\farcs1$ (semi-axes) region. The right column shows the spectra extracted from the same region but before spatially integrating the emission in each channel, the emission in each pixel is shifted in velocity to account for the projected Keplerian velocity at that location in the disk \citep{Yen2016, Teague_gofish}. This effectively removes the Keplerian line broadening from the line profile, narrowing the line and increasing the peak signal-to-noise ratio. This is particularly evident for the \ce{C^17O} spectrum whereas the \ce{N2H+} spectrum does not improve after the line stacking possibly due to the coarser spectral resolution of the data and the hyperfine components. 

\begin{figure}
\centering
\includegraphics[width=\hsize]{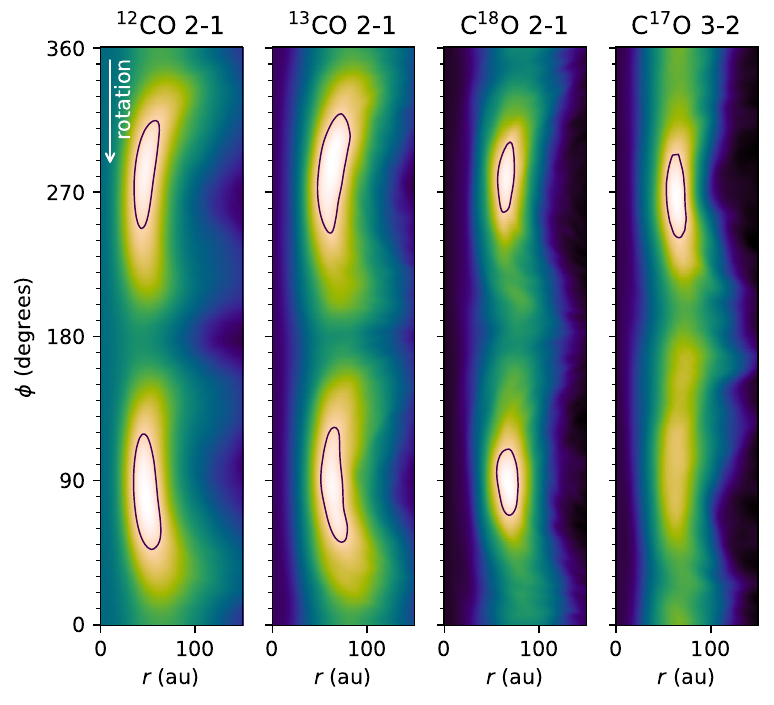}
  \caption{Polar deprojection of the emission in the PDS~70 disk starting from in the minor axis on the north east (far) side of the disk in anticlockwise direction. The three left most panels present the \ce{^12CO}, \ce{^13CO}, and \ce{C^18O} $J=2-1$ emission imaged with a robust parameter of 2.0 from \citet{Rampinelli2024}. The fourth panel shows the \ce{C^17O} $J=3-2$ line emission presented in this work imaged with a robust parameter of 2.0. We note that the beam size for all four images is not circular in the frame of the disk. The black contour is set to 90\% of the peak emission in each panel. Only the \ce{C^17O} emission shows a tentative asymmetry.}
     \label{fig:polar_COiso}
\end{figure}

\begin{figure}
\centering
\includegraphics[width=\hsize]{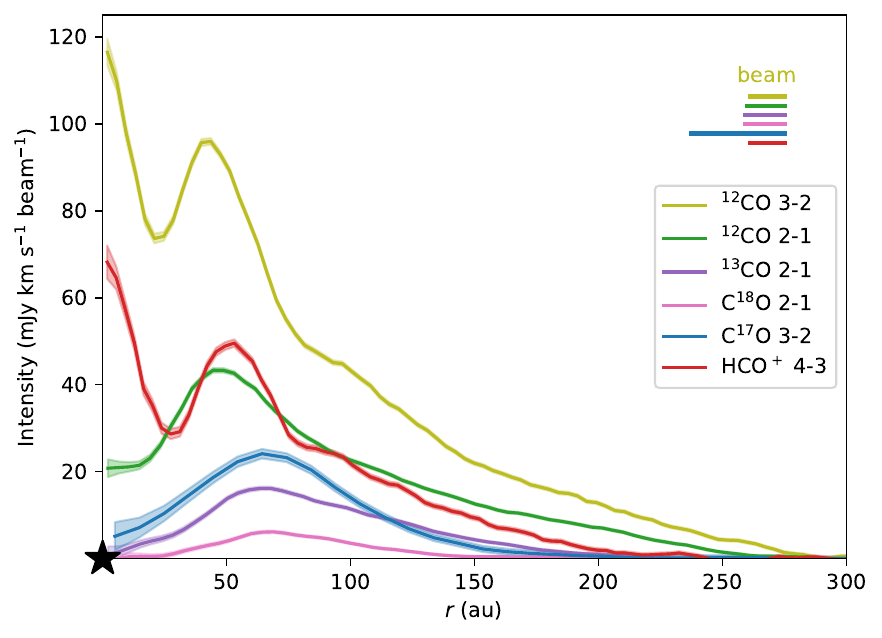}
  \caption{Azimuthally averaged radial profiles of archival \ce{^12CO} $3-2$ and $2-1$, \ce{^13CO} $2-1$, \ce{C^18O} $2-1$, \ce{C^17O} $3-2$, and \ce{HCO+} $4-3$ emission in this disk \citep{Facchini2021, Rampinelli2024}. }
     \label{fig:azi_avg_all}
\end{figure}

A comparison between the \ce{C^17O} data and archival CO isotopologue and \ce{HCO+} data in this disk are presented in Fig.~\ref{fig:polar_COiso} and Fig.~\ref{fig:azi_avg_all}. The former shows the polar deprojection of the \ce{^12CO}, \ce{^13CO}, and \ce{C^18O} $J=2-1$ emission is compared to the that of the \ce{C^17O} $J=3-2$ emission. Only the \ce{C^17O} emission shows a clear asymmetry along the azimuth of the disk. The latter figure presents the azimuthally averaged radial profiles of the \ce{^12CO} $J=3-2$, $J=2-1$, \ce{^13CO} $J=2-1$, \ce{C^18O} $ J=2-1$, \ce{C^17O} $J=3-2$, and the \ce{HCO+} $J=4-3$ transitions \citep{Facchini2021, Rampinelli2024}. The bright \ce{^12CO} $J=3-2$ and \ce{HCO+} $J=4-3$ emission show that PDS~70 disk hosts a gas gap rather than a gas cavity fully cleared of gas.

\begin{table*}
\caption{Summary of the observations taken as part of ALMA project 2022.1.01477.S covering the \ce{N2H+} $1-0$ transition in Band~3 and the \ce{C^17O} $3-2$ transition in Band~7. \label{tab:obs_details}}
\centering
\begin{tabular}{cccccccccc}
\hline\hline
molecule & Band & EB & date & baselines & integration  & pwv  & bandpass \& flux  & phase  & number of \\
& & & &  (m) & time$^{(a)}$ (min) & (mm) & calibrator & calibrator & antennas  \\
\hline
\ce{N2H+}  & 3 & SB EB0 & 26-03-2023 & 15-1400 & 43 & 5.0 & J1427-4206 & J1407-4302 & 46 \\
\ce{N2H+}  & 3 & LB EB0 & 02-07-2023 & 85-8500 & 45 & 0.7 & J1427-4206 & J1407-4302 & 42 \\
\ce{N2H+}  & 3 & LB EB1 & 02-07-2023 & 85-8500 & 45 & 0.6 & J1427-4206 & J1407-4302 & 43 \\
\ce{C^17O} & 7 & SB EB0 & 23-01-2023 & 15-783  & 31 & 0.4 & J1427-4206 & J1352-4412 & 47 \\
\ce{C^17O} & 7 & SB EB1 & 23-01-2023 & 15-783  & 31 & 0.4 & J1427-4206 & J1352-4412 & 46 \\
\ce{C^17O} & 7 & LB EB0 & 23-05-2023 & 79-3638 & 40 & 0.9 & J1427-4206 & J1352-4412 & 43 \\
\ce{C^17O} & 7 & LB EB1 & 01-06-2023 & 27-3638 & 40 & 0.5 & J1427-4206 & J1352-4412 & 42 \\
\ce{C^17O} & 7 & LB EB2 & 02-06-2023 & 27-3638 & 40 & 0.5 & J1427-4206 & J1352-4412 & 42 \\
\ce{C^17O} & 7 & LB EB3 & 03-06-2023 & 27-3638 & 40 & 0.7 & J1427-4206 & J1352-4412 & 40 \\
\ce{C^17O} & 7 & LB EB4 & 04-06-2023 & 27-3638 & 40 & 0.6 & J1427-4206 & J1352-4412 & 40 \\
\ce{C^17O} & 7 & LB EB5 & 07-06-2023 & 77-3638 & 40 & 0.6 & J1427-4206 & J1352-4412 & 43 \\
\ce{C^17O} & 7 & LB EB6 & 08-06-2023 & 27-3638 & 40 & 0.7 & J1427-4206 & J1352-4412 & 43 \\
\ce{C^17O} & 7 & LB EB7 & 27-11-2023 & 31-3638 & 40 & 0.7 & J1427-4206 & J1352-4412 & 44 \\
\ce{C^17O} & 7 & LB EB8 & 28-11-2023 & 31-3638 & 40 & 0.8 & J1427-4206 & J1352-4412 & 45 \\
\hline
\end{tabular}
\tablefoot{$^{(a)}$ Science time on the PDS~70 disk. }
\end{table*}

\section{\texttt{mcmc} autocorrelation time} \label{app:mcmc}

The default length of each \texttt{mcmc} sampling is 3000 steps with an acceptance fraction of $\sim$~30\%. The acceptance fraction varies between 22\% for the \texttt{mcmc} sampling that only fits to \ce{C^18O}, \ce{N2H+}, and the 273 GHz continuum and 33\% where the mcmc fits to all observed line and continuum fluxes and an excitation temperature of 11 K is used to rescale the \ce{N2H+} emission. 

\begin{figure*}
\centering
\includegraphics[width=\hsize]{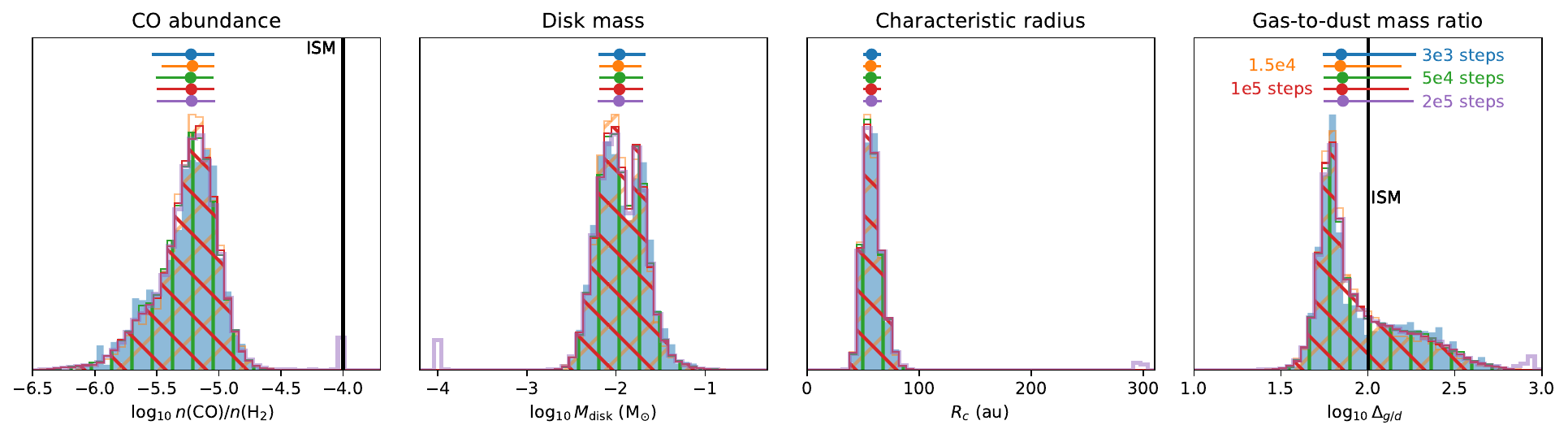}
  \caption{Posterior distribution for the peak CO abundance, disk mass, $R_{\mathrm{c}}$, and the gas-to-dust mass ratio for an \texttt{mcmc} run with $3\times 10^3$ steps (blue; fiducial; same as the blue histograms in Fig.~\ref{fig:mcmc}), $1.5\times 10^4$ steps (orange hatched), $5\times 10^4$ steps (green hatched), $1\times 10^5$ steps (red hatched), and $2\times 10^5$ steps (purple).}
     \label{fig:mcmc_Nsteps}
\end{figure*}

\begin{figure}
\centering
\includegraphics[width=\hsize]{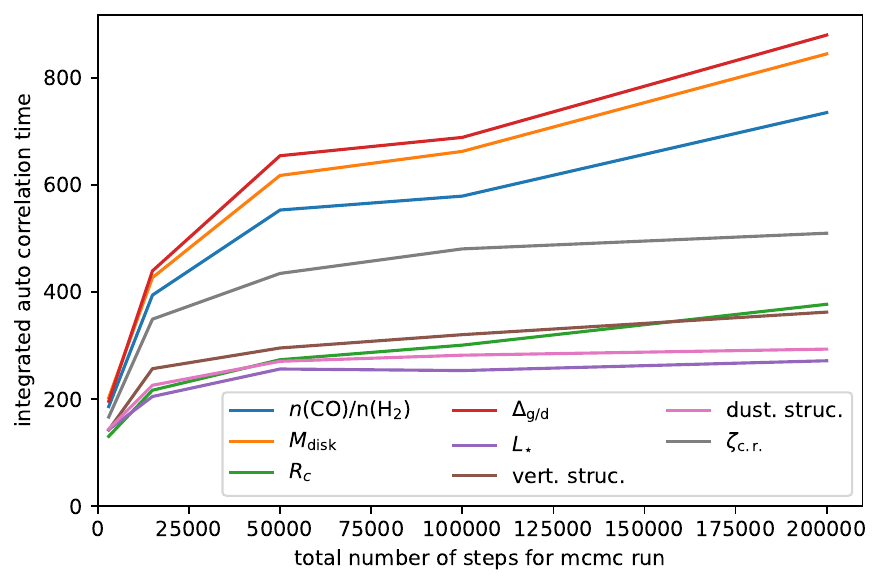}
  \caption{Integrated autocorrelation times for the posteriors presented in Fig.~\ref{fig:mcmc_Nsteps}. }
     \label{fig:mcmc_autocorr}
\end{figure}

Figure~\ref{fig:mcmc_Nsteps} demonstrates that the posterior distributions of the run including \ce{C^17O} do not change substantially if the \texttt{mcmc} sampling is run for a larger amount of steps up to $2\times 10^5$. The integrated autocorrelation time for each parameter are presented in Fig.~\ref{fig:mcmc_autocorr}. The integrated autocorrelation time corresponding to the unconstrained parameters (dust structure, vertical structure, and cosmic ray ionization rate) and to the stellar luminosity (characterized by an approximately Gaussian posterior) flattens off after 200-500 steps. The integrated autocorrelation time for the CO abundance, disk mass, characteristic radius, and the gas-to-dust mass ratio keep increasing. This is likely due to the more complex structure of the posteriors that are bimodal and/or show a heavy tail. The increase in their autocorrelation times between \texttt{mcmc} samplings using $1\times 10^5$ steps and $2\times 10^5$ steps could be due to the small additional peak that appears at $n(\rm{CO})/n(\rm{H_2}) = 10^{-4}$, $M_{\rm disk} = 10^{-4}$~M$_{\odot}$, $R_c = 300$~au, and $\Delta_{\rm g/d} = 10^3$. These peaks coincide with the edge of the underlying allowed parameter space and are likely not representative of the PDS~70 disk. This is supported by the fact that the \ce{^12CO} $J=2-1$ and $J=3-2$ transitions are not seen outside 300~au (see Fig.~\ref{fig:azi_avg_all}, \citealt{Facchini2021, Rampinelli2024}). Furthermore, these peaks are sufficiently small that they do not change the derived disk parameters. Therefore, 3000 steps are sufficient to sample the posterior distribution.

\section{Scaling the \ce{N2H+} line flux} \label{app:Tex}

We detect the \ce{N2H+} $1-0$ transition in the PDS~70 disk. However, this specific transition is not included in the modelling grid of \citet{Trapman2025_exoALMA, Trapman2025_agepro}. Therefore, this flux is scaled to the $3-2$ transition that is included in the modelling grid as discussed in Sect.~\ref{sec:methods_mcmc}. In this subsection, we explore the effect of the assumed excitation temperature on the derived disk parameters. 

Based on thermochemical modelling in disks \ce{N2H+} is expected to be abundant between the CO and \ce{N2} snow surfaces and in the region where \ce{N2} is self-shielding but CO is not \citep{vantHoff2017}. Therefore, the bulk of the \ce{N2H+} reservoir is expected to emit from just below the CO snowsurface at $\sim 20$~K. This is consistent with the findings in the TW~Hya disk where both \ce{N2H+} $1-0$ and $4-3$ are detected \citep{Schwarz2019}. Table~\ref{tab:N2H+_vs_T} summarizes the derived \ce{N2H+} flux for three different excitation temperatures using Eq.~\ref{eq:scaling_N2H+} and assuming optically thin emission. If the emission is optically thick, the line flux scales directly with the frequencies of the lines as $(f_{3-2}/f_{1-0})^2$ to a value of $0.68 \pm 0.13$~Jy~km~s$^{-1}$. In principle, the hyperfine components of the \ce{N2H+} $1-0$ line could be used to constrain the optical depth. However, the spectral resolution of the data at $\sim 3.1$~km~s$^{-1}$ is too coarse to separate the hyperfine components at $\sim -7$~km~s$^{-1}$ and $\sim 8$~km~s$^{-1}$ from the main component. 

\begin{figure*}
\centering
\includegraphics[width=\hsize]{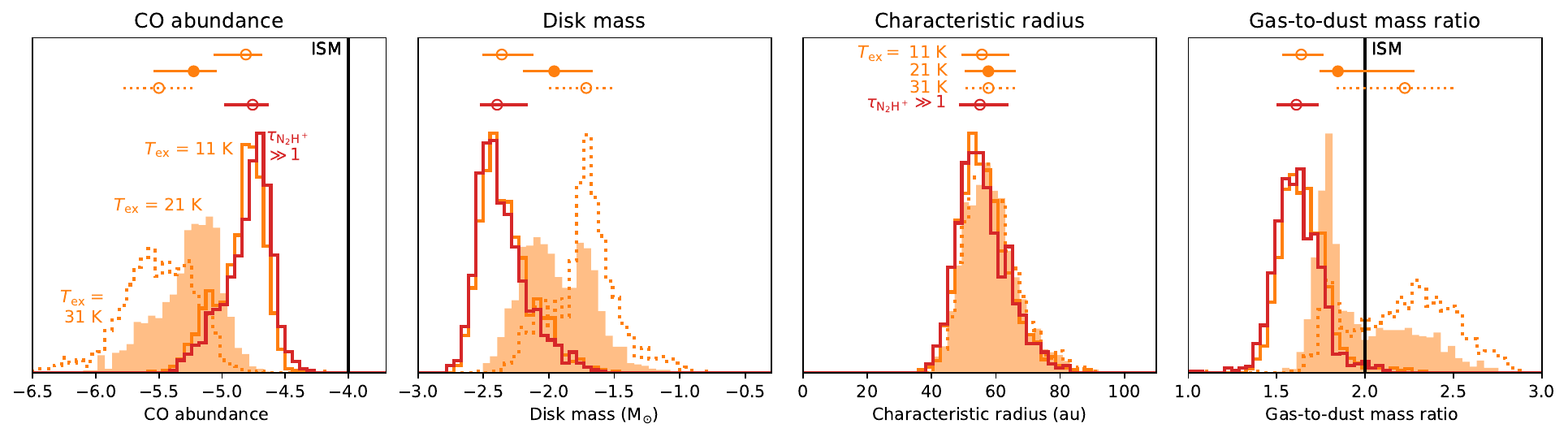}
  \caption{Posterior distribution for the peak CO abundance, disk mass, $R_{\mathrm{c}}$, and the gas-to-dust mass ratio for different scalings of the \ce{N2H+} $1-0$ to the $3-2$ line flux. All orange histograms assume optically thin emission. The filled orange histograms are identical to those in Fig.~\ref{fig:mcmc}, the solid, open orange histogram and the dashed open histogram show the same but then for an \ce{N2H+} $3-2$ flux obtained using an excitation temperature of 11 and 31~K, respectively. The red histogram presents the posteriors assuming the \ce{N2H+} flux is optically thick. }
     \label{fig:mcmc_N2H+}
\end{figure*}

The resulting posterior distributions for the various scalings of the \ce{N2H+} flux are presented in Fig.~\ref{fig:mcmc_N2H+}. The three orange histograms show the results assuming that the \ce{N2H+} emission is optically thin with an excitation temperature of 11~K (solid open histograms), 21~K (solid filled histograms; fiducial), and 31~K (dotted open histograms). The red histogram, on the other hand, shows the posteriors when the \ce{N2H+} emission is assumed to be optically thick. The derived peak CO abundance increases by a factor of $\sim 2.5$ when the \ce{N2H+} emits from a colder layer with a temperature of 11~K or is optically thick compared to the fiducial assumption of optically thin emission with $T_{\rm ex} = 21$~K. This shift is driven by the lower \ce{N2H+} $3-2$ flux in this case. The higher peak CO abundance then results in a lower derived disk mass and a somewhat lower gas-to-dust mass ratio though the posterior distributions overlap with those of the fiducial case. If the \ce{N2H+} emits from a warmer layer at 31~K and is optically thin, the derived disk parameters are consistent within the uncertainty of the \texttt{mcmc} sampling procedure. All in all, CO is sequestered compared to the ISM abundance independent of the assumed \ce{N2H+} excitation temperature and assumption on optically thin or optically thick emission.

\begin{table}[!bp]
\caption{\ce{N2H+} $3-2$ line flux for various assumed excitation temperatures assuming optically thin emission. \label{tab:N2H+_vs_T}}
\centering
\begin{tabular}{cc}
\hline\hline
Assumed $T_{\rm ex}$ (K) & $F_{\nu}\Delta V$ (Jy~km~s$^{-1}$)\\
\hline
11 & $0.80 \pm 0.15$ \\
21$^{(a)}$ & $2.12 \pm 0.40$  \\
31 & $2.98 \pm 0.57$  \\
\hline
\end{tabular}
\tablefoot{$^{(a)}$ Fiducial value for the fitting. }
\end{table}

\section{Line fluxes predicted by the \texttt{mcmc} chains} \label{sec:mcmc_fluxes}

\begin{figure*}
\centering
\includegraphics[width=\hsize]{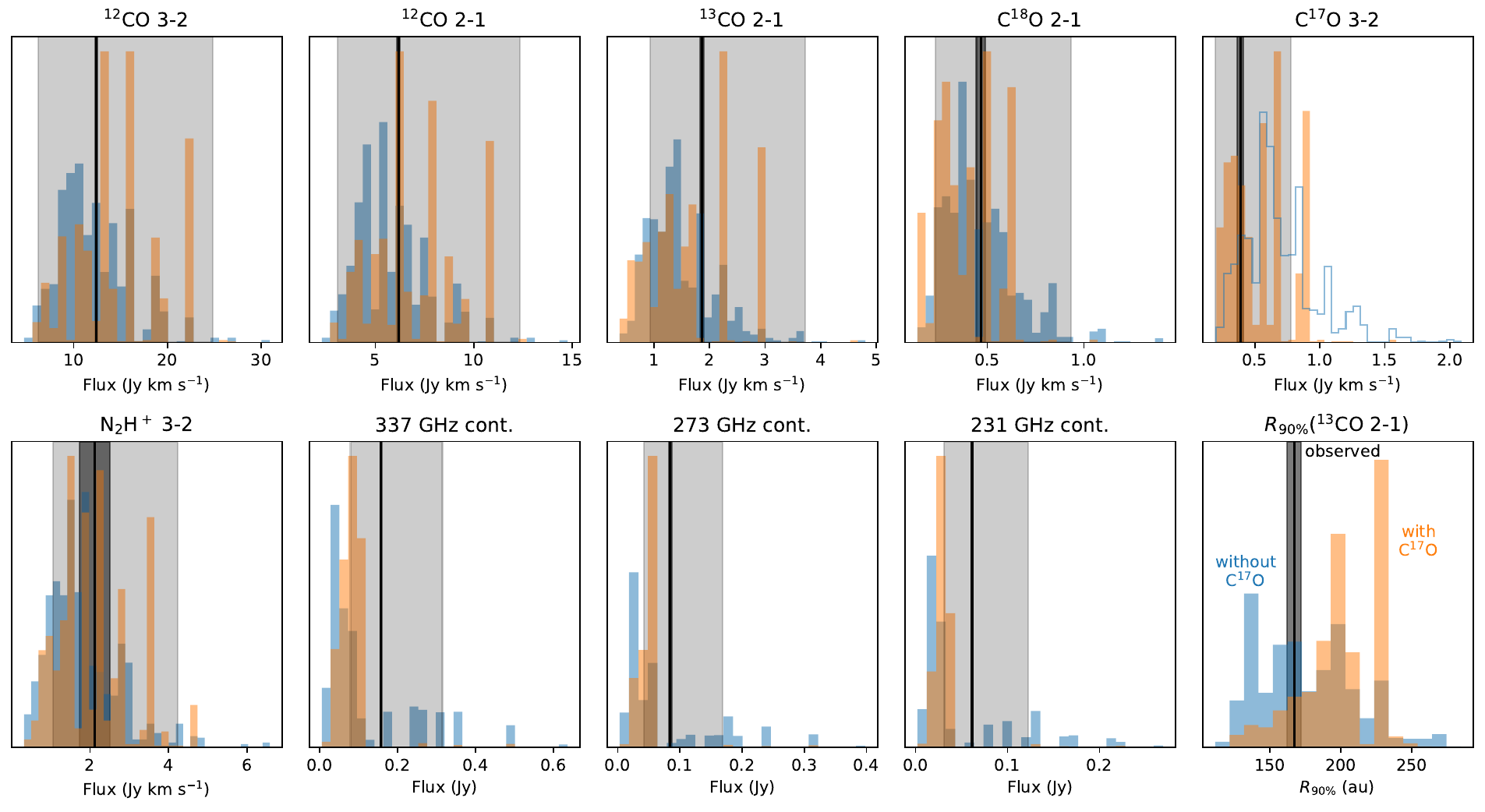}
  \caption{Line and continuum fluxes and 90\% light radius of the \ce{^13CO} $2-1$ line predicted by the \texttt{mcmc} chains with (orange) and without (blue) fitting for the \ce{C^17O} flux used to derive the posteriors in Fig.~\ref{fig:mcmc}. The blue histogram for the \ce{C^17O} (top right panel) is shown as an open histogram as \ce{C^17O} is not included in the \texttt{mcmc} sampling. The vertical black lines indicate the observed values with their associated uncertainties (shaded dark grey region). The light grey shaded region for the line fluxes show the typical model uncertainty of a factor of 2 (not included in the \texttt{mcmc} sampling; \citealt{Bruderer2012, Bruderer2013}). }
     \label{fig:mcmc_fluxes}
\end{figure*}

Figure~\ref{fig:mcmc_fluxes} presents the line and continuum fluxes as well as the 90\% light radius of the \ce{^13CO} $2-1$ emission predicted by the models along the \texttt{mcmc} chains of Fig.~\ref{fig:mcmc}. The distributions for the line fluxes are similar between the \texttt{mcmc} sampling with (orange) and without (blue) fitting for the \ce{C^17O} emission and they are generally centered around or close to the observed values (black vertical lines). The distribution of the predicted $R_{\rm 90\%}$ shows a much broader distribution than the observed value in both cases. Interestingly, the range of predicted \ce{C^17O} fluxes largely overlaps for both samplings. This shows that the models including \ce{C^17O} do not fail at reproducing the \ce{C^17O} line fluxes. However, including \ce{C^17O} seems to induce a reweighting of the family of models resulting in a shift in the predicted disk parameters.

\section{Effect of $R_{\rm 90\%}$(\ce{^13CO})} \label{app:R90}

\begin{figure}
\centering
\includegraphics[width=\hsize]{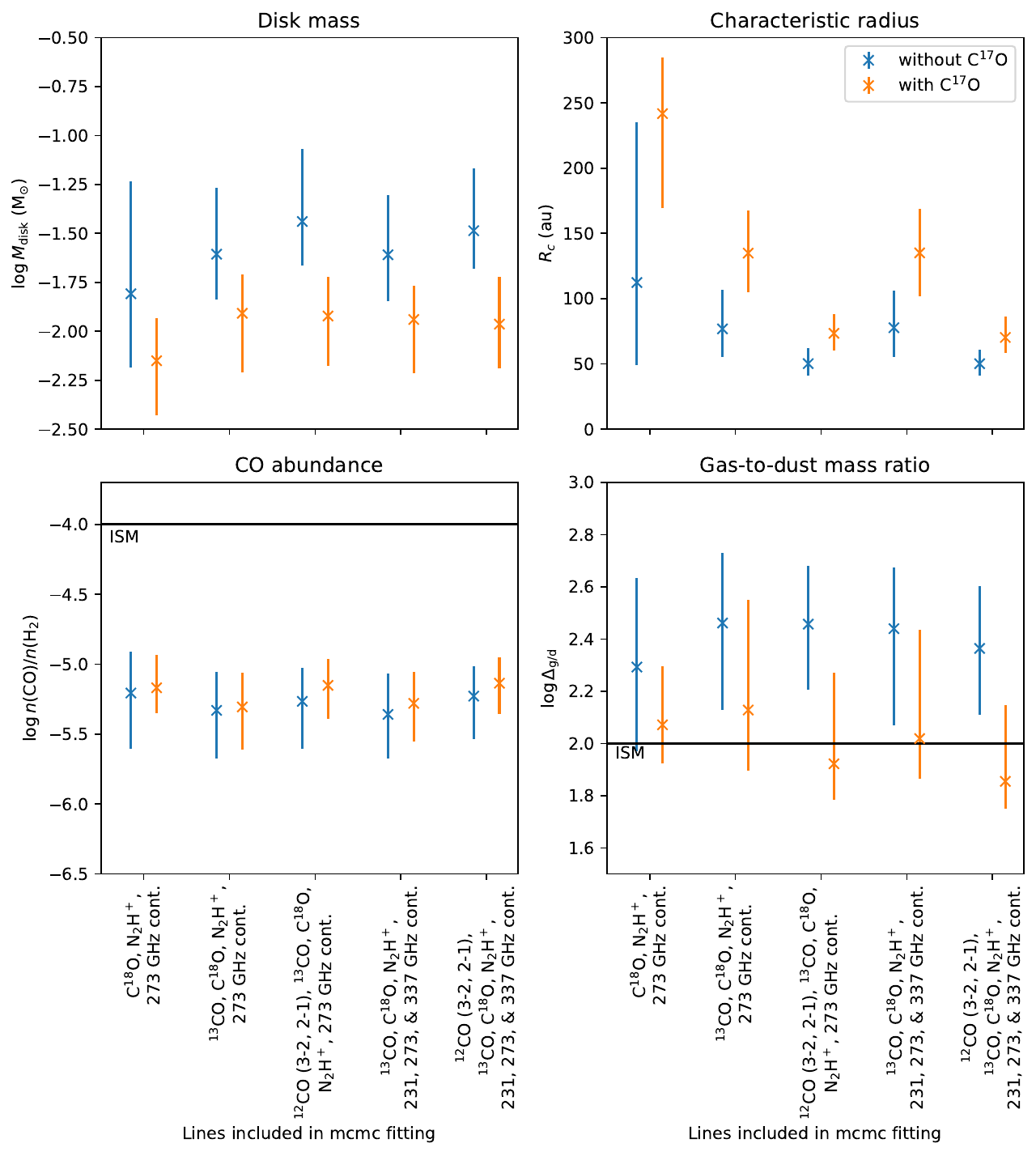}
  \caption{Same as Fig.~\ref{fig:fluxes_to_fit} but then not fitting for the 90\% flux radius of the \ce{^13CO} $J=2-1$ emission. Note that the y-axis for the top right panel is different than in Fig.~\ref{fig:fluxes_to_fit}.}
     \label{fig:fluxes_to_fit_noR90}
\end{figure}

\begin{figure*}
\centering
\includegraphics[width=\hsize]{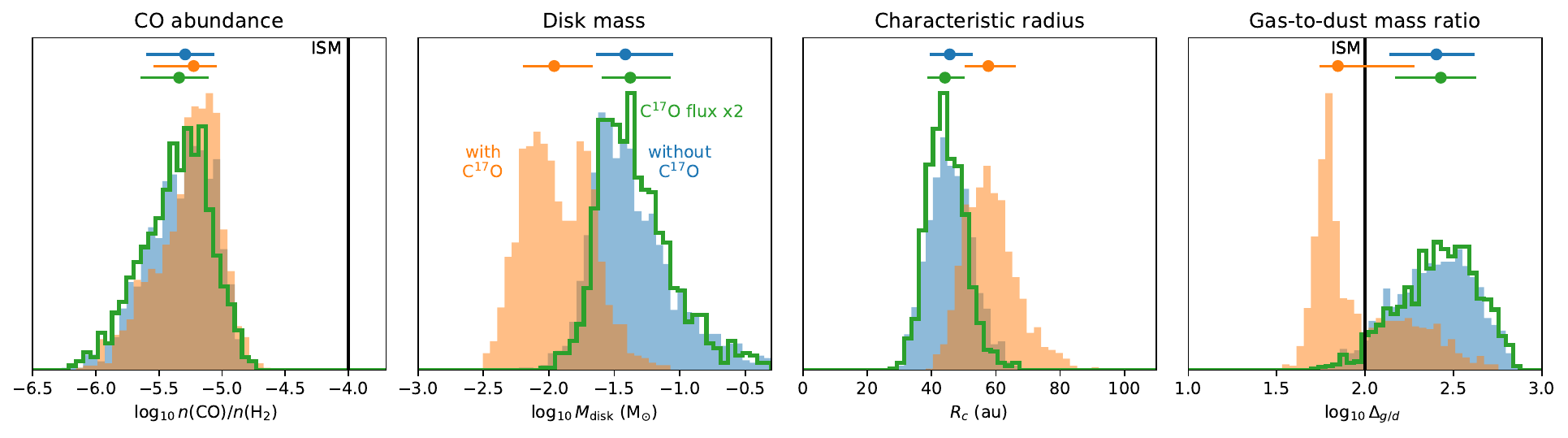}
  \caption{Posterior distribution for the peak CO abundance, disk mass, $R_{\mathrm{c}}$, and the gas-to-dust mass ratio. The blue and orange histograms show the posteriors when excluding (blue) and including (orange) the \ce{C^17O} flux in the \texttt{mcmc} sampling. The green histogram shows the posteriors when only the \ce{C^17O} flux is doubled artificially. We note that these green histograms are very similar to the blue ones.}
     \label{fig:mcmc_C17Ox2}
\end{figure*}

The \texttt{mcmc} sampling in the main text all include the 90\% flux radius of the \ce{^13CO} $J=2-1$ emission as \citet{Stapper2022} demonstrated that including the \ce{^12CO} radius in the fitting can help to break some of the degeneracies in the fitting. We omitted the 90\% flux radius of the \ce{^13CO} $2-1$ for all \texttt{mcmc} fits presented in Sect.~\ref{sec:fluxes_to_fit} and present the results in Fig.~\ref{fig:fluxes_to_fit_noR90}. The most striking difference is that $R_{\mathrm{c}}$ is much less constrained when $R_{\rm 90\%}$(\ce{^13CO}) is omitted from the fitting. For all \texttt{mcmc} runs sampling for at least the \ce{^13CO}, \ce{C^18O}, \ce{N2H+}, and the 273~GHz continuum emission, the $R_{\rm 90\%}$(\ce{^13CO}) does not affect the derived disk mass, peak CO abundance, and gas-to-dust mass ratio within the 16$^{\rm th}$-84$^{\rm th}$ percentiles of the posterior distributions. The dichotomy in the disk mass due to whether the \ce{C^17O} line flux is included in the fitting persists. Only when the \ce{C^17O} flux is increased by a factor of 2, this dichotomy is resolved (see Fig.~\ref{fig:mcmc_C17Ox2}).

\section{Bootstrapping without fitting for \ce{C^17O}}
\begin{figure*}
\centering
\includegraphics[width=\hsize]{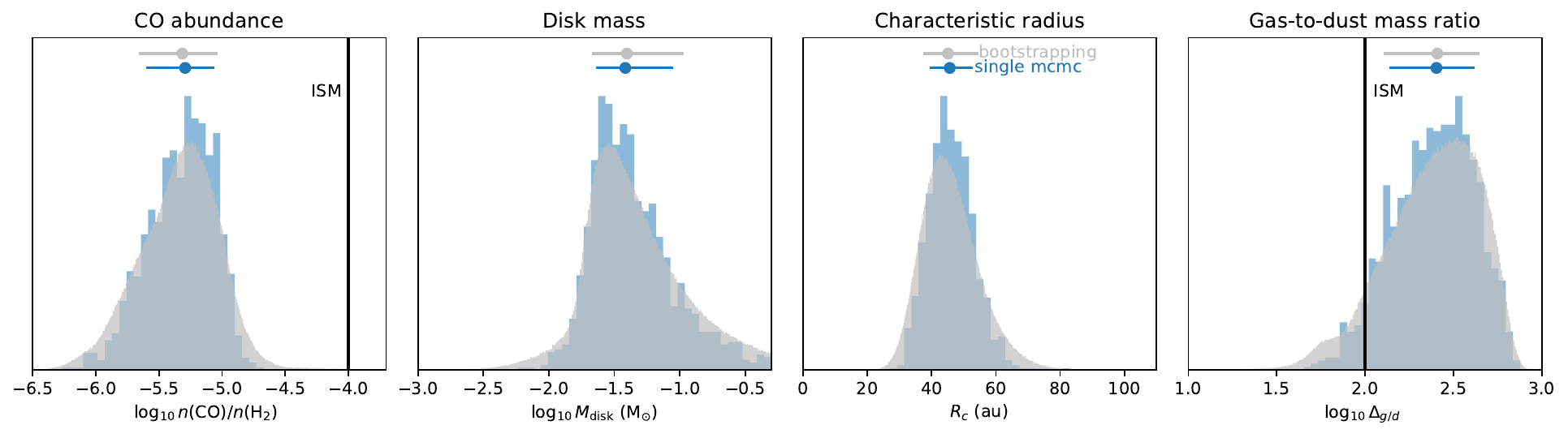}
  \caption{Distribution of the posteriors for derived peak CO abundance, disk mass, $R_{\mathrm{c}}$, and gas-to-dust mass ratio for 1000 \texttt{mcmc} runs with input fluxes and $R_{\rm 90\%}$(\ce{^13CO}) values sampled from the measured fluxes (grey). The \ce{C^17O} flux is not fitted for in the \texttt{mcmc} sampling. The blue histograms are identical to those in Fig.~\ref{fig:mcmc} and show the result of a single \texttt{mcmc} fit using the measured values and their associated uncertainties.}
     \label{fig:bootstrapping_noC17O}
\end{figure*}

The results of the bootstrapping fitting for $R_{\rm 90\%}$(\ce{^13CO}) and all fluxes in Table~\ref{tab:fit_inputs} except for the \ce{C^17O} line flux are presented in Fig.~\ref{fig:bootstrapping_noC17O}. Similar to the discussion in Sect.~\ref{sec:bootstrapping}, the posterior distribution is comparable to that of a single \texttt{mcmc} run fitting to truly measured fluxes, indicating that the uncertainty on the derived parameters is not dominated by the uncertainty on the measured fluxes but rather by the underlying modelling grid.

\end{document}